\documentclass[sn-mathphys-num]{sn-jnl}

\usepackage{natbib}
\usepackage{graphicx}%
\usepackage{multirow}%
\usepackage{amsmath,amssymb,amsfonts}%
\usepackage{amsthm}%
\usepackage{mathrsfs}%
\usepackage[title]{appendix}%
\usepackage[dvipsnames]{xcolor}%
\usepackage{textcomp}%
\usepackage{manyfoot}%
\usepackage{booktabs}%
\usepackage{algorithm}%
\usepackage{algorithmicx}%
\usepackage{algpseudocode}%
\usepackage{listings}%

\definecolor{RoyalBlue}{RGB}{65, 117, 186}
\definecolor{ForestGreen}{RGB}{33, 141, 33}
\definecolor{lavenderindigo}{RGB}{148, 90, 235}
\definecolor{brightcerulean}{RGB}{28, 171, 215}
\definecolor{ultramarine}{RGB}{0,32,96}

\begin{document}

\title[Article Title]{Multi-controlled single-qubit unitary gates based on the quantum Fourier transform and deep decomposition}

\author*[1]{\fnm{Vladimir V.} \sur{Arsoski}}\email{vladimir.arsoski@etf.bg.ac.rs}

\affil*[1]{\orgdiv{The Department of Microelectronics and Technical Physics}, \orgname{School of Electrical Engineering - University of Belgrade}, \orgaddress{\street{Bulevar kralja Aleksandra 73}, \city{Belgrade}, \postcode{P.O. Box 35–54}, \country{Serbia}}}

\abstract{We will present a few new generalizations of the multi-controlled X (MCX) gate that uses the quantum Fourier transform (QFT). Firstly, we will optimize QFT-MCX and prove that it is equivalent to a stair MCX gates array. This stair-wise structure will allow us to devise a method for adding an arbitrary phase factor to each qubit. The first MCX generalization into multi-controlled unitary gates (MCU) relies on replacing phase gates acting on the target qubit with controlled unitary gates. We will employ alternative single-qubit gate notation to minimize the complexities of these gates and show how to expand the circuit straightforwardly to the multi-controlled multi-target (MCMT) gate. The second generalization relies on the ZYZ-like decomposition. We will show that by extending one QFT-MCX circuit we implement the two multi-controlled X gates needed for the decomposition. Finally, we will split control wirelines into groups and use iterative ZYZ-like decomposition on QFT-MCU to obtain ``deep decomposed'' (DD) MCU which employs a lower number of C-NOTs than the previous two, thus making DD-MCU less prone to decoherence and noise. The supremacy of our implementations over the best-known optimized algorithm will be demonstrated by emulating noisy quantum calculations.}

\keywords{Quantum computing, Quantum algorithms, Multi-controlled gates, Multi-target gates, Quantum Fourier transform, Optimization}

\pacs[MSC Classification]{03G12, 81P68}

\maketitle

\section{Introduction}\label{sec1}

Current quantum devices are constrained by the number of qubits available and the noise introduced during the execution of nonideal quantum operations employing the native gates used for computation in a genuine quantum device. The hardware of these Noisy Intermediate-Scale Quantum devices (NISQ) \cite{Preskill2018} is constantly improving by increasing the number of qubits and fidelity of the native gates used in a particular quantum computing architecture. However, the efficiency in performing quantum computation can be improved by optimizing software that defines a quantum circuit implementation. It can be achieved by more efficient error mitigation \cite{Kim2023}, quantum state preparation \cite{Plesch2011, Zhang2021, Araujo2021}, and decomposition of unitary gates into scalable quantum circuits in the basis set of elementary gates \cite{Barenco1995, Nielsen2010, Shende2006, Malvetti2021}. The decomposition of a circuit is usually not unique. Therefore, different optimization techniques can be used to minimize the circuit's depth and (or) the number of elementary gates used \cite{Bae2020, Brugiere2021, Cuomo2023}.

The first algorithm for decomposing multi-controlled $U(2)$ gates, which doesn't use auxiliary qubits, was proposed in Ref.~\cite{Barenco1995}. It exhibits a quadratic increase in the circuit depth and number of elementary gates with the number of control qubits. The authors showed that implemented circuits can be efficiently reduced by removing some gates at the price of phase relativization, using ancilla qubits, and approximating gates up to the target error $\epsilon$. The advantages of using relative-phase Toffoli gates to obtain linear depths of $n$-qubit MCX gates were first recognized in Ref.~\cite{Saeedi2013}. The effectiveness of this approach is demonstrated theoretically \cite{Maslov2016} and experimentally \cite{Jun2023}. To correct phases, an additional $(C-R_x)$-based circuit is introduced in Ref.~\cite{Silva2022} that approximately doubles the complexity of the previously simplified circuit. Different approximations are used to reduce the number of elementary gates in this circuit \cite{Vale2023, Silva2023}. As for all other cases, using auxiliary qubits lowers the circuit depth \cite{He2017, Balauca2022, Arsoski2024}.

After this brief introduction, we will preview the previously implemented QFT-based MCX circuits and various methods for their optimization in sections \ref{sec2}. We will introduce equivalent circuits to simplify MCX and prove that QFT-MCX building blocks are equivalent to a stair-wise cascade of MCX gates in section \ref{sec3}. Furthermore, using deductive analysis of phase correction in the QFT-MCX, we will devise a method to add different phase factors to each qubit, thus making our circuit usable for other more complex applications. The detailed analysis of three QFT-based MCU implementations is presented in section \ref{sec4}. The first implementation is obtained by modifying QFT-MCX's controlled gates acting on the target qubit. We introduce a general single-qubit rotation to minimize the most complex gates and show how to generalize our circuit to the multi-target gate. The second implementation uses the ZYZ-like decomposition, but only one modified QFT-MCX to implement two MCX operations needed in the decomposition, which makes it twofold simpler than the standard one. Our implementations utilize the devized phase correction method to control the desired phase factor applied to the target qubit. Finally, the third implementation employs dividing control qubits into clusters and recursive ZYZ-like decomposition to minimize the number of C-NOT gates used in MCU making it less susceptible to noise. In section \ref{sec5}, a significant advantage over the state-of-the-art implementation is demonstrated by transpiling circuits and emulating noisy quantum computations. The most important conclusions are summarized in the section \ref{sec6}. We prove the equivalence of the state-of-the-art circuit with our QFT-MCU in Appendix \ref{sec7}.

\section{Related work}\label{sec2}

Our approach is based on the QFT and inspired by basic quantum arithmetic \cite{Draper2000}. One may show that simple  QFT-based increment/decrement by one can be used to implement multi-controlled X gates. The equivalence between the QFT-based MCX and the standard one is rigorously proved theoretically in Ref.~\cite{Arsoski2024}. However, a fairly simple explanation can give an insight into the principles of the proposed implementation.

\begin{figure}[ht]
  \centering
  \includegraphics[scale=0.8]{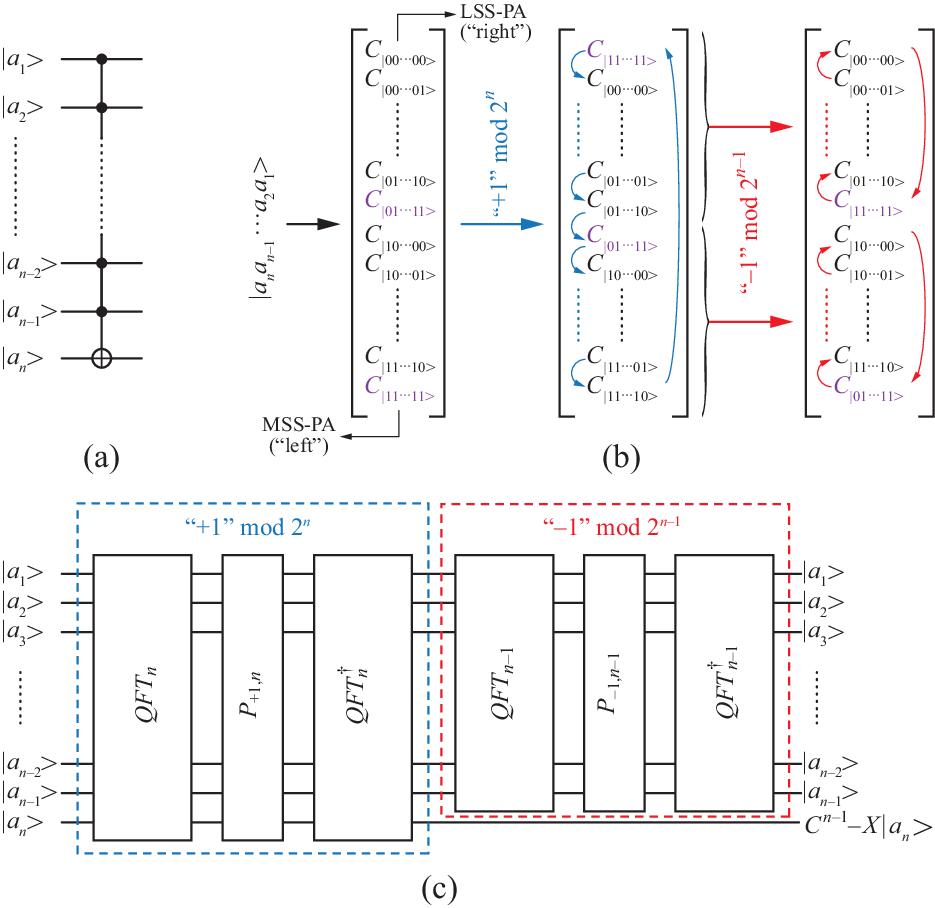}
  \caption{(a) Schematic view of $n$-qubit multi-controlled $X$ gate, (b) illustration of increment/decrements action on the state vector, and (c) implementations block diagram using QFTs and phase gates, adapted from Fig.1 of  Ref.~\cite{Arsoski2024} originally published in Quantum Information Processing by Springer Nature. In panel (b), MSS-PA (most significant state's probability amplitude) in classical computing ordering is on the left, while LSS-PA (least significant state's probability amplitude) is on the right.}
  \label{fig1}
\end{figure}

Multi-controlled gate, shown in Fig.~\ref{fig1}(a), executes $X$ operation on the highest $n{\rm th}$ qubit based on the state of $(n-1)$ lower qubits. A pure $n$-qubit state $\vert a\rangle=\vert a_na_{n-1}\cdots a_2a_1\rangle$ is represented by a $2^n$ dimensional state vector in Hilbert space. The set of orthonormal basis states $\{ \vert 000\cdots000\rangle, \vert 000\cdots001\rangle,\cdots,\vert 111\cdots110\rangle,\vert 111\cdots111\rangle \}=\{\vert0\rangle,\vert1\rangle,\cdots,\vert2^n-2\rangle, \vert2^n-1\rangle\}$ span this linear vector space. Each term in the state vector $C_{\vert k\rangle}$ (where $k\in\{0,2^n-1\}$) is the probability amplitude (weight) of the corresponding basis state. An application of MCX on a $n$-qubit system results in the swap between $C_{\vert 011\cdots 111\rangle}$ and $C_{\vert 111\cdots 111\rangle}$ weights. 

Arithmetic operations performed on a $n$-qubit state are congruent modulo $2^n$. To implement MCX, we initially increment a qubit state by one. The weight of $\vert k\rangle$ state becomes the weight of $\vert (k+1) \mod 2^n\rangle$. Thus, all weights are circularly shifted by one, as shown in Fig.~\ref{fig1}(b). Performed on a classical register that stores bits, this operation is known as the circular shift to the left. To restore values of control qubits, we have to execute decrement by one on the quantum register comprised of lower $(n-1)$ qubits. Thus, we perform the quantum circular shift to the right in the basis subsets $\vert 0 b_{n-1}\cdots b_m\cdots b_1\rangle$ and $\vert 1 b_{n-1}\cdots b_m\cdots b_1\rangle$ where $b_m\in\{ 0,1\}$, as displayed in Fig.~\ref{fig1}(b). As a result, we swapped target states' probability amplitudes thereby implementing the MCX operation. A block schematic of $n$-qubit MCX implementation is shown in Fig.~\ref{fig1}(c).

A standard QFT implementation uses phase gates:
\begin{equation}
	R_m=P\left(\frac{\pi}{2^{m-1}}\right)=\begin{bmatrix} 1 & 0 \\ 0 & e^{\i\frac{2\pi}{2^m}} \end{bmatrix}=\begin{bmatrix} 1^{1/2^{m-1}} & 0 \\ 0 & (e^{i\pi})^{1/2^{m-1}} \end{bmatrix}=Z^{1/2^{m-1}}.\label{eqRm}
\end{equation}
Increments and decrements by one can be executed using a QFT-based adder \cite{Draper2000}. This method computes QFT on the first addend and uses $R_m$ gates to evolve it into QFT of the sum based on the second addend. Then, the inverse of the QFT (QFT$^\dagger$) returns the result to the computational basis. Thus, the increment by one is executed by $QFT^\dagger_n P_{+1,n} QFT_n\vert a\rangle=\vert a + 1 \mod 2^n \rangle=C^{n-1}X\vert a_n\rangle\otimes \vert a_{n_c}+1 \mod 2^{n-1}\rangle$, where $P_{+1,n}=Z^{1/2^{n-1}}\otimes Z^{1/2^{n-2}} \otimes\cdots\otimes Z^{1/2}\otimes Z$ and $\vert a_{n_c}\rangle=\vert a_{n-1}a_{n-2}\cdots a_2a_1\rangle$ is the register comprised of lower $(n-1)$ control qubits. This part of the schematic is framed by a {\color{RoyalBlue}blue} dashed line and labeled by {\color{RoyalBlue}``$+1$''$\mod 2^n$} in Fig.~\ref{fig1}(c). To restore control qubits to the initial value, we execute decrement by one $ I_{2\times 2} \otimes QFT^\dagger_{n-1} P_{-1,n-1} QFT_{n-1}(C^{n-1}X\vert a_n\rangle\otimes \vert a_{n_c}+1 \mod 2^{n-1}\rangle)=C^{n-1}X\vert a_n\rangle\otimes\vert a_{n_c} \rangle$, where $P_{-1,n-1}={Z^{1/2^{n-2}}}^\dagger\otimes \cdots\otimes {Z^{1/2}}^\dagger\otimes Z^\dagger$ and $I_{2\times 2}$ is the identity $2\times 2$ matrix. This part of the circuit is framed by a {\color{red}red} dashed line and labeled by {\color{red}``$-1$''$\mod 2^{n-1}$} in Fig.~\ref{fig1}(c).

\begin{figure}[ht]
  \centering
  \includegraphics[scale=0.63]{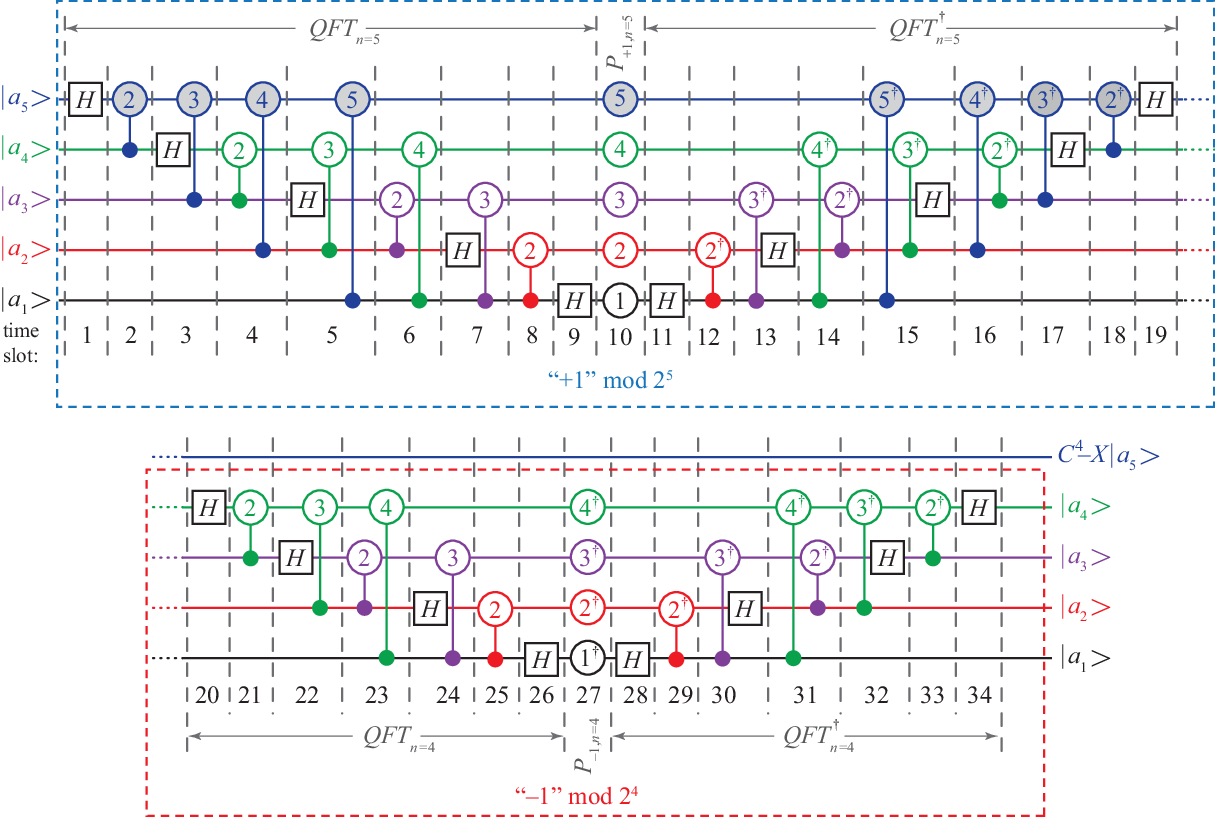}
  \caption{The decomposition of the 5-qubit MCX using non-elementary $H$, $R_m$ and $C-R_m$ gates. The circle indicates phase gate $R_m$ with index $m$ inscribed. Gates are divided into time slots by vertical gray dashed lines, where the index of the time slot is denoted at the bottom.}
  \label{fig2}
\end{figure}

The decomposition of the 5-qubit QFT-MCX is shown in Fig.~\ref{fig2}. To estimate the time complexity of the circuit, gates that can execute simultaneously are assembled in a single time slot bounded by a vertical dashed gray line in Fig.~\ref{fig2}. Control qubits are $\vert a_1\rangle - \vert a_4 \rangle$, and the target is $\vert a_5\rangle$. Let us consider the {\color{RoyalBlue}``$+1$''$\mod 2^5$} part of the circuit and find the expression for the operation that executes on the target qubit. Controlled phase gates ($C-R_m=C-Z^{1/2^{m-1}}$), which act on the target qubit, are represented by gray circles. There are $(2n-1)$ of these gates. The first $(n-1)$ gates (to the left) are conditioned on the qubits of the control register $\vert a_c\rangle=\vert a_{n-1}\cdots a_1\rangle$, the central gate is unconditional, and the last $(n-1)$ gates are controlled by $\vert a_c+1 \mod 2^{n-1}\rangle=\vert a_{n-1}^\prime\cdots a_1^\prime\rangle$. In a simplified notation, the action of these gates on the target wireline is
\begin{align}
	&\left({Z^{\frac{a_{n-1}^\prime}{2^1}}}^\dagger {Z^{\frac{a_{n-2}^\prime}{2^2}}}^\dagger \cdots {Z^{\frac{a_{1}^\prime}{2^{n-1}}}}^\dagger\right) {Z^{\frac{1}{2^{n-1}}}} \left( Z^\frac{a_1}{2^{n-1}} Z^\frac{a_2}{2^{n-2}}\cdots Z^\frac{a_{n-1}}{2}\right)\nonumber\\
&=\left( Z^{\frac{2^0 a_1^\prime+ 2^1 a_2^\prime +\cdots 2^{n-3} a_{n-2}^\prime + 2^{n-2} a_{n-1}^\prime}{2^{n-1}}} \right)^\dagger Z^{\frac{1+2^0 a_1+ 2^1 a_2 +\cdots 2^{n-3} a_{n-2} + 2^{n-2} a_{n-1}}{2^{n-1}}}\nonumber\\
&={Z^\frac{1+a_c\mod 2^{n-1}}{2^{n-1}}}^\dagger Z^\frac{1+a_c}{2^{n-1}}.\label{eqZ}
\end{align} 
This operator is the identity ($I_{2\times 2}$) in all cases except for $a_1=a_2=\cdots=a_{n-2}=a_{n-1}=1$ when $(1+a_c\mod 2^{n-1})=0$ and $1+a_c=2^{n-1}$ so the expression, given by Eq.~\ref{eqZ}, is $Z$. Since $HZH=X$, when the controls are uncomputed by {\color{red}``$-1$''$\mod 2^{n-1}$}, the overall circuit executes MCX. 

\section{QFT-MCX circuit optimization and a phase implementation}\label{sec3}

\begin{figure}[ht]
  \centering
  \includegraphics[scale=0.55]{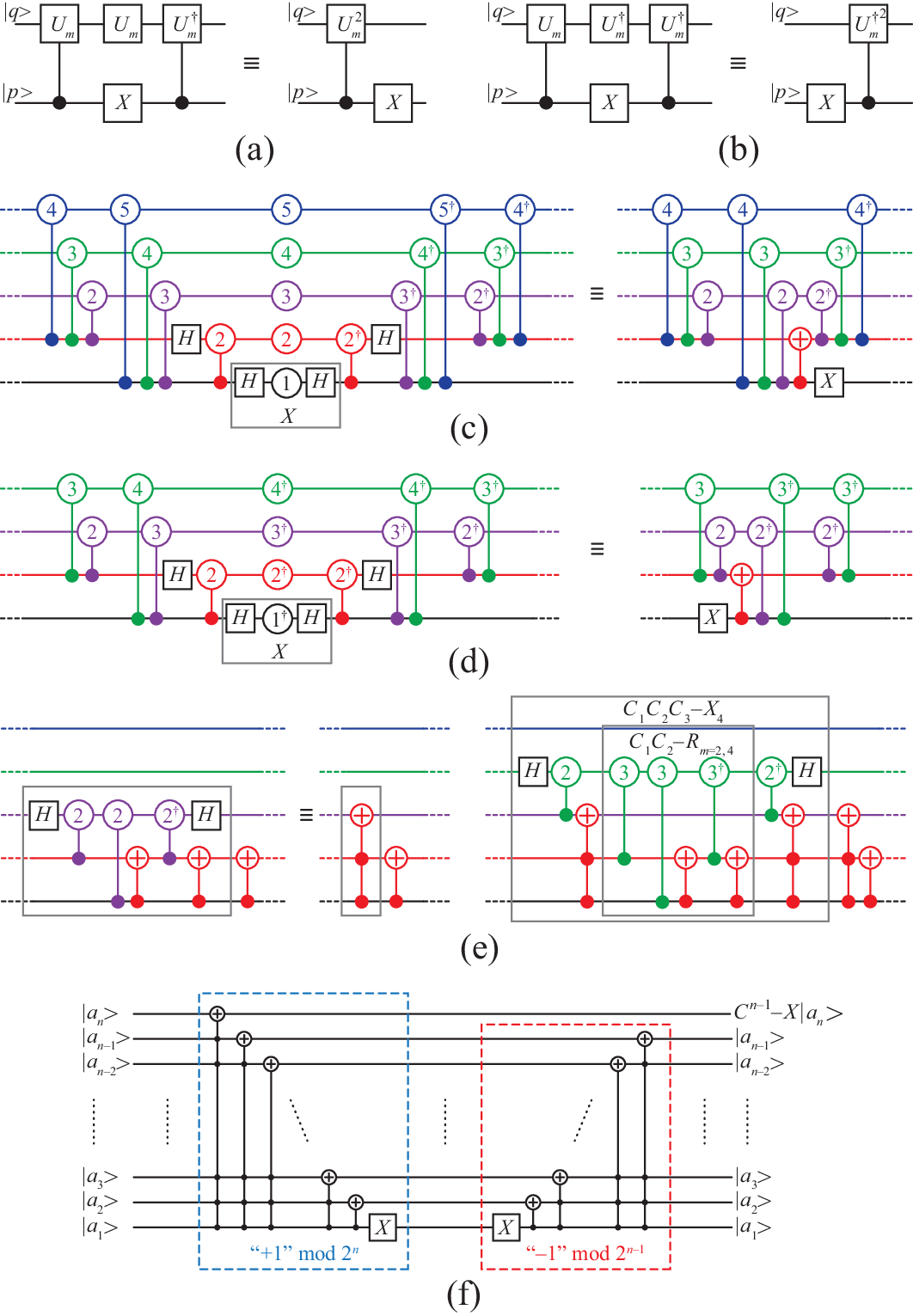}
  \caption{The identities for merging gates in (a) {\color{RoyalBlue}``$+1$''$\:{\rm mod}\: 2^n$} and (b) {\color{red}``$-1$''$\:{\rm mod} \:2^{n-1}$}, where $U_m$ is a single qubit unitary gate (Note: for $U=R_m=Z^{1/2^{m-1}}$, $U^2=R_{m-1}=Z^{1/2^{m-2}}$). Merging of phase gates and controlled phase gates for the 5-qubit MCX in (c) {\color{RoyalBlue}``$+1$''$\:{\rm mod}\: 2^5$} and (d) {\color{red}``$-1$''$\:{\rm mod} \:2^4$} circuits, respectively. (e) The basic schematics illustrate how our circuit is equivalent to a series of MCX gates and (f) the equivalent stair-wise schematic of QFT-MCX based on Eq.~\ref{eqStairs}.}
  \label{fig3}
\end{figure}

Let's analyze some circuit simplifications. On the first wireline there is $HR_1H=HZH=X$ (slots 9 to 11 in Fig.~\ref{fig2}) and $HR_1^\dagger H=HZ^\dagger H = X^\dagger = X$ (slots 26 to 28 in Fig.~\ref{fig2}). The phase gates controlled by the first qubit are applied at the beginning of QFT and end of QFT$^\dagger$. These gates can merge using the equivalence shown in Figs.~\ref{fig3}(a) and (b), respectively. We conclude there is no need to implement $P_{\pm1}$ separately. Moreover, the QFT in {\color{RoyalBlue}``$+1$''$\mod 2^n$} and the QFT$^\dagger$ in {\color{red}``$-1$''$\mod 2^{n-1}$} will also have a set of $C-R_m$ (or $C-R_m^\dagger$) gates less. Optimized parts of {\color{RoyalBlue}``$+1$''$\:{\rm mod}\: 2^n$} and {\color{red}``$-1$''$\:{\rm mod} \:2^{n-1}$} circuits are displayed in Figs.~\ref{fig3}(c) and (d), respectively. It is straightforward to show that the optimization reduces the number of time slots in MCX by 8. Although slots comprising only a few gates are merged, it can still be significant when applying MCX on a small set of qubits. Non-optimized QFT-based MCX becomes less complex than the standard implementation when the number of qubits $n>6$ \cite{Arsoski2024}. Using the above optimization, one may show that QFT-MCX is better or just as good as the optimized standard implementation even for $n\leq6$.

Next, we will explain the functionality of {\color{RoyalBlue}``$+1$''$\:{\rm mod}\: 2^n$} and {\color{red}``$-1$''$\:{\rm mod} \:2^{n-1}$} blocks. Using an iterative approach, we show that the {\color{RoyalBlue}``$+1$''$\mod 2^n$} circuit
\begin{eqnarray}
	&&QFT^\dagger_n P_{+1,n} QFT_n\vert a\rangle=\vert a + 1 \mod 2^n \rangle \nonumber \\
	&=&(C_1C_2\cdots C_{n-2}C_{n-1}-X\vert a_n\rangle)\otimes \vert a_{n_c}+1 \mod 2^{n-1}\rangle \nonumber \\
	&=& (C_1\cdots C_{n-1}-X\vert a_n\rangle)\otimes(C_1\cdots C_{n-2}-X\vert a_{n-1}\rangle)\otimes \vert a_{n_c-1}+1 \mod 2^{n-2}\rangle \nonumber \\
	&=&(C_1\cdots C_{n-1}-X\vert a_n\rangle)\otimes\cdots\otimes (C_1-X\vert a_2\rangle)\otimes(X\vert a_1\rangle),\label{eqStairs}
\end{eqnarray}
is equivalent to the stair-wise array of controlled $X$ gates, and {\color{red}``$-1$''$\mod 2^n$} is its inverse. Alternatively, by adding to $C-X$ gates on the first two wirelines and using Lemma 6.1 from Ref.~\cite{Barenco1995} it is evident that two $C-X$ and three $C-R_2$ gates implement Toffoli gate, as shown in Fig.~\ref{fig3}(e). Applying this equivalence on $C-R_3$ gates and recurrently on higher qubits, we obtain an equivalent $C_1C_2C_3-X_4$ gate, as shown in Fig.~\ref{fig3}(e). From this perspective, the schematic view of the QFT-MCX circuit is given in Fig.~\ref{fig2}(f).

\begin{figure}[h!]
  \centering
  \includegraphics[scale=0.72]{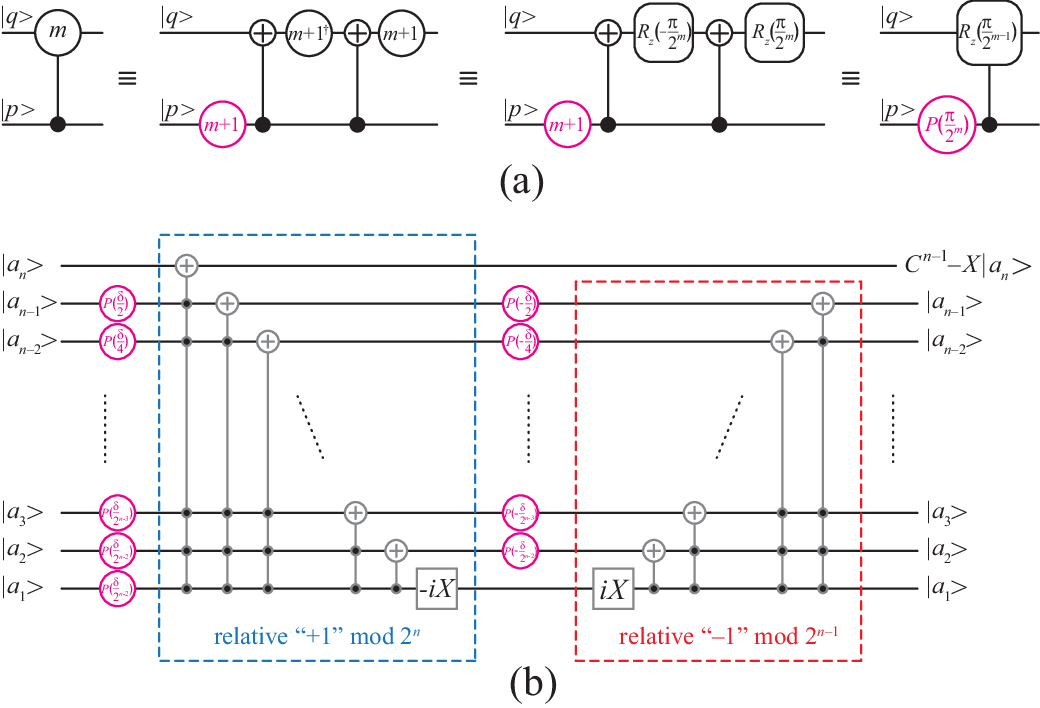}
  \caption{(a) Decompositions of $C-R_m$ gates using {\color{magenta}$R_{m+1}=P(\frac{\pi}{2^m})$} (marked by circles) and $R_z(\pm \frac{\pi}{2^m})$ gates (marked with rounded squares). Here, $R_z(\pm \frac{\pi}{2^m})$ (or $R_{m+1}$) gate acting on $\vert q\rangle$ can be shifted to the beginning of the wireline. (b) A schematic of the simplified QFT-MCX obtained by omitting the phase correction gates $P$ from $C-R_m$ acting on control wirelines. A gray color of multi-controlled gates indicates a relative phase $-i$ and $+i$ on gates in the {\color{RoyalBlue}``$+1$''} and {\color{red}``$-1$''} blocks, respectively.}
  \label{fig4}
\end{figure}

Different forms of $C-R_m$ decompositions are schematically displayed in Fig.~\ref{fig4}(a). The control phase gate $C_p-R_{m,q}$ applied to $\vert a_p\rangle$ as the control and $\vert a_q\rangle$ ($q>p$ in Fig.~\ref{fig2}) as the target is
\begin{eqnarray}
	&&(I_{2\times 2}\otimes P(\frac{\pi}{2^m}))(C-X)(I_{2\times 2}\otimes P(-\frac{\pi}{2^m}))(C-X)(P(\frac{\pi}{2^m})\otimes I_{2\times 2})(\vert a_p \rangle \otimes \vert a_q\rangle) \nonumber\\
	&=&(I_{2\times 2}\otimes R_z(\frac{\pi}{2^m}))(C-X)(I_{2\times 2}\otimes R_z(-\frac{\pi}{2^m}))(C-X)(P(\frac{\pi}{2^m})\otimes I_{2\times 2})(\vert a_p \rangle \otimes \vert a_q\rangle)\nonumber \\
	&=&(C-R_z(\frac{\pi}{2^{m-1}}))({\color{magenta}P(\frac{\pi}{2^m})}\otimes I_{2\times 2})(\vert a_p \rangle \otimes \vert a_q\rangle),\label{eqRz}
\end{eqnarray}
where we used the identity $P(\gamma)XP(-\gamma)=R_z(\gamma)XR_z(-\gamma)$. 

A phase difference between $C-R_m$ and $C-R_z(\frac{\pi}{2^{m-1}})$ gates is compensated using the ${\color{magenta}R_{m+1}=P(\frac{\pi}{2^{m}})}$ gate on the control wireline of the $C-R_z$ gate, as explained in Ref.~\cite{Barenco1995}. If we neglect the phase-adding $P$ gates and use $C-R_z$ instead of $C-R_m$, providing that all qubits lower than the selected one are in the state $\vert 1\rangle$, $R_z(\pi)=-iZ$ will be implemented between the two Hadamard gates in {\color{RoyalBlue}``$+1$''$\:{\rm mod}\: 2^n$} instead of $Z$ (see Fig.~\ref{fig2} and Eq.~\ref{eqZ}). The omitted $P$ gates implement the phase factor $i$, which can be easily verified in Fig.~\ref{fig2}. Consequently, there is a phase relativization in the operation implemented on each wireline of the {\color{RoyalBlue} relative ``$+1$''$\:{\rm mod}\: 2^n$} subcircuit. Thus, on the $k$th wireline we will implement a multi-controlled $(-iX)$ gate conditioned by the lower $(k-1)$ qubits. Similarly, multi-controlled $(iX)$ gates are implemented by the {\color{red}relative ``$-1$''$\:{\rm mod} \:2^{n-1}$} subcircuit. Since $-i\cdot i=1$, it will not affect the control qubits outputs. Moreover, based on the symmetry argument and using the decomposition of $C-R_m$ from Fig.~\ref{fig4}(a), it is straightforward to show that on each control wireline, displayed in Fig.~\ref{fig2}, these $P$ gates cancel out if all lower control qubits are one (when we have $P(\frac{\pi}{2^m})XP(-\frac{\pi}{2^m})P(\frac{\pi}{2^m})XP(-\frac{\pi}{2^m})=I_{2\times 2}$), or if they are not (when we have $P(\frac{\pi}{2^m})P(-\frac{\pi}{2^m})P(\frac{\pi}{2^m})P(-\frac{\pi}{2^m})=I_{2\times 2}$). However, this doesn't apply to the $C-R_m$ gates acting on the target wireline (see gray colored circles in Fig.~\ref{fig2}), where there is no ``inverted'' sequence of gates in {\color{red}``$-1$''} block-circuit (which would perform ``uncomputation''). Therefore, the ${\color{magenta}P}$ gates (shown in {\color{magenta}magenta} in Fig.~\ref{fig4}(b)) emerging in the $C-R_z$ decomposition of the $C-R_m$ gates acting on the target wireline can not be omitted. In Fig.~\ref{fig4}(b), we show an equivalent schematic of QFT-MCX based on the relative phase multi-controlled $(\mp iX)$ gates comprised of $C-R_z$s and additional ${\color{magenta}P}$ gates originated from $C-R_m$ acting on the target qubit. We used the notation $P(\delta / 2^{m-1})$ instead of $R_m=P(\pi / 2^{m-1})$,  to consider a general case of adjusting the phase-factor ($e^{i\delta}$) applied to the target qubit.

If all controls are $\vert 1\rangle$, $-iX$ is implemented to the target qubit. To correct the additional phase ($-i$), we need to implement the phase factor $e^{i\delta}=e^{i\pi/2}=i$ to the target qubit by adding $P(\delta/2)$ and $P(-\delta/2)$ to the $(n-1)$th wireline, as shown in Fig.~\ref{fig4}(b), thereby implementing
\begin{equation}
	X P\left(-\frac{\delta}{2} \right) X P\left(\frac{\delta}{2}\right) = R_z(\delta) = e^{-i\delta/2}P(\delta). \label{eqPh}
\end{equation}
By doing this, the phase factor $e^{-i\delta/2}$ is introduced on the $(n-1)$th wireline conditioned on the lower $(n-2)$ qubits, which can be corrected by adding $P(\delta/4)$ and $P(-\delta/4)$ to the $(n-2)$th wireline, as shown in Fig.~\ref{fig4}(b). However, this results in the additional phase $e^{-i\delta/4}$ (the wireline output is $e^{-i\delta/4}P(\delta/2)$, by analogy to Eq.~\ref{eqPh}), which can be corrected by adding appropriate phase gates to the lower qubit, and so on. Phase gates employed for the conditional application of $e^{i\delta}$ to the target qubit are displayed in {\color{magenta}magenta} in Fig.~\ref{fig4}(b). One may note that these gates annihilate if one of the control qubits is in $\vert 0\rangle$ state. Moreover, the same ``phase-adding'' circuit can be implemented by surrounding the {\color{red}``$-1$''$\:{\rm mod} \:2^{n-1}$} subcircuit with the same phase gates, but in reverse ordering to the one used in the {\color{RoyalBlue}``$+1$''$\:{\rm mod}\: 2^n$}. Thereby, on the $(n-1)$th wireline we implement
\begin{equation}
	P\left(\frac{\delta}{2} \right) X P\left(-\frac{\delta}{2}\right) X = R_z(\delta) = e^{-i\delta/2}P(\delta),
\end{equation}
providing that all controls are $\vert 1\rangle$. Alternatively, we may combine these two approaches. 

The essence of phase correction is to use two multi-controlled X gates existing in this stair-wise structure to implement a multi-controlled $R_z$ gate on each wireline. So, on the $(n-1)$th wireline we implement $C^{(n-2)}-R_z(\delta)$, on the $(n-2)$th $C^{(n-3)}-R_z(\delta/2)$, and on the first wireline we apply $R_z(\delta/2^{n-2})$. One should note that in genuine quantum computing, we utilize $R_z$ gates instead of {\color{magenta}$P$} resulting in the global phase $e^{i\delta/2^{n-1}}$ to the output state. Choosing the symmetric configuration may avoid this at the price of keeping the two consecutive $X$ gates on the first wireline.

Moreover, we can conditionally apply a custom phase to each control qubit by choosing different arguments in the ${\color{magenta}P}$ gates sequence. This is potentially very useful from the point of view of a more general application that combines multiple circuits. For example, we can use less complex MC-SU(2) gates, which are linear in the depth and the number of C-NOTs used. After performing the set of SU(2)-based computations, we can custom-add phase factors for each qubit using only one circuit which, as we will show in the next chapter, employs a quadratic number of C-NOT gates.

\begin{figure}[h!]
  \centering
  \includegraphics[scale=0.68]{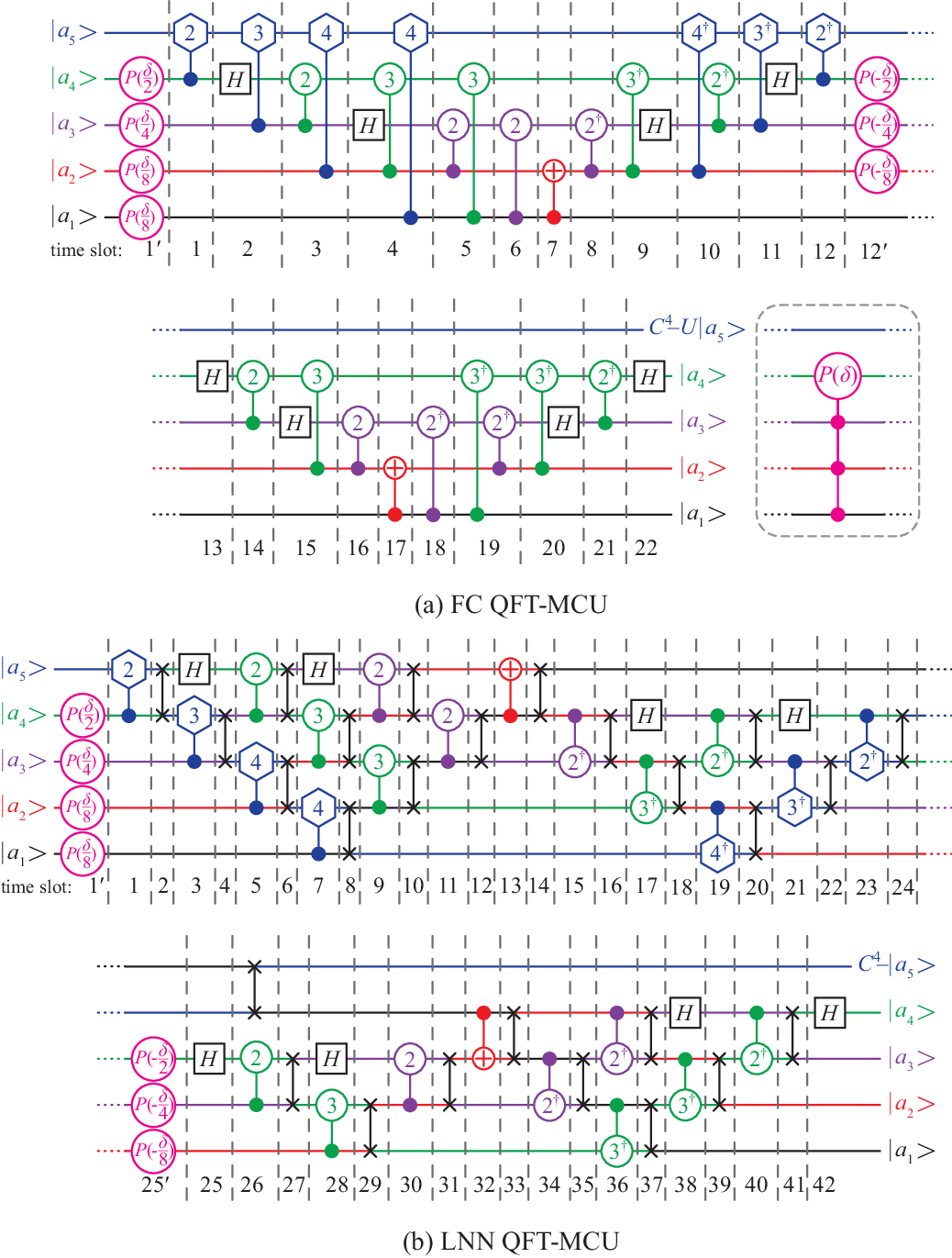}
  \caption{A schematic of five-qubit MCU implementations based on QFT-MCX modification in the (a) fully connected (FC) and (b) linear nearest-neighbor (LNN) quantum computer architecture. Circles denote either phase $R_m=Z^{1/2^{m-1}}$ or $R_z(\pi/2^{m-1})$ gates, while hexagons represent special unitary ${SU}_m={SU}^{1/2^{m-1}}$ gates ($U_m=e^{i\delta/2^{m-1}}{SU}_m$), with the value $m$ inscribed into gate symbols. To estimate the time complexity, gates that can execute simultaneously are divided into time slots. We do not assign a number for the time slot comprising {\color{magenta}$P$} gates since they can merge or execute in parallel with some other.}
  \label{fig5}
\end{figure}

\section{Multi-controlled U(2) gates}\label{sec4}

The aim is to generalize the MCX circuit to the multi-controlled single-qubit unitary gate (MCU). We will explore implementations in two distinct quantum computer architectures to find the lower and upper bounds for the time and space complexities. The most favorable architecture supports interaction between arbitrary pairs of qubits, implying that the system is fully connected (FC). Implementation in this architecture is the least complex, setting the lower complexity bounds. The most restricted is the linear one allowing interactions only between nearest-neighbours (LNN). Using this architecture demands swapping many qubits to perform two-qubit gates, and additional SWAP gates increase the time and space complexities. We use two strategies to generalize the circuit. The first is based on modifying the QFT-based MCX gate, while the second uses extended optimized QFT-MCX gates to implement a circuit similar to the complex gate that could be obtained using the well-known ZYZ ($ABC$) decomposition \cite{Barenco1995}.

It is straightforward to conclude from Eq.~\ref{eqZ} that by substituting $R_m=Z^{1/2^{m-1}}$ with $U_m=U^{1/2^{m-1}}$ on the target qubit and omitting $H$ gates, the circuit implements a multi-controlled unitary $U$ gate. Schematic views of 5-qubit MCUs in the FC and LNN architecture are given in Figs.~\ref{fig5}(a) and (b), respectively. If $U$ is not a special unitary, then $U(2)=e^{i\delta}SU(2)$, where $\delta$ is a real valued. 
To implement an arbitrary multi-controlled $U(2)\not\in SU(2)$, we need to implement multi-controlled $SU(2)$ and the controlled phase circuit $C_1C_2\cdots C_{n-2}-P_{n-1}(\delta)$ that adds the phase factor $e^{i\delta}$, as shown in the inset in Fig.~\ref{fig5}(a). In section \ref{sec3} we found that the {\color{RoyalBlue}``$+1$''} and {\color{red}``$-1$''} blocks are equivalent to a sequence of multi-controlled gates needed to implement $(n-1)$-controlled phase gate. So, we don't need to implement it using a separate circuit. Moreover, these gates can be found by direct decomposition of controlled-$U(2)$ gate
\begin{equation}
	C-U_m= (C-{SU}_m)(P(\delta/2^{m-1})\otimes I_{2\times 2}),
\end{equation}
where $U_m=e^{i\delta/2^{m-1}}{SU}_m$. Furthermore, these phase gates execute simultaneously with a gate acting on the controlled wireline. Therefore, we have not assigned them a separate time slot in Fig.~\ref{fig5}.

One may show that FC QFT-MCU executes in $(8n-18)$ time slots. It comprises of $4(n-3)$ single-qubit Hadamards, $2(n-1)(n-3)$ controlled phase gates, two $C-$NOTs and $(2n-3)$ $C-U_m$ gates. An approximate QFT (AQFT) may provide greater accuracy than a full QFT in the presence of decoherence \cite{Barenco1996}. Controlled phases $R_m$ with $m\leq[\log_2 n]$ are used in AQFT. Therefore, the number of controlled gates reduces to $2([\log_2 n]-1)(2n-3-[\log_2n])$ $C-R_m$ gates and at least $2([\log_2n]-1)$ $C-U_m$ gates.

Using the systematic approach for swapping qubits in finite-neighbor quantum architectures \cite{Maslov2007}, one may find that LNN QFT-MCU needs $(8n-20)$ time slots with $(2n^2-6n+6)$ SWAP gates. Therefore, LNN uses approximately twice the number of time slots and gates as FC QFT-MCU, thus setting the upper bound for the time and space complexities. In the LNN architecture, SWAP and $C-R_m$ are neighboring. Decomposing $C-R_m$ and $C-R_m^\dagger$ into $C-R_z(\pm\pi/2^{m-1})$ gate and shifting $R_z(\mp\pi/2^{m})$ either to the beginning or the end of controlled gate sequence, two $C-X$ gates cancel out between SWAP and $C-R_z$. Hence, each additional SWAP gate (except the upper gate in slot 26) effectively reduces to a C-NOT gate, while the target and control of the $C-X$ gate in the neighbor $C-R_z$ swap over.

Actual circuit depth and the number of gates implemented depend on the native gate set (NGS) of the quantum device used in the calculation. The decomposition of a non-elementary gate uses a few elementary gates executed in a certain number of time slices. This number of time slices defines the circuit depth. One should note that some elementary gates can be executed in parallel and some cancel out, as explained in Ref.~\cite{Arsoski2024} for the selected NGS. Elementary gates that comprise the NGS have a high but finite fidelity. Also, current quantum devices are prone to noise and decoherence. Therefore, low-depth circuits using fewer elementary gates are less prone to errors.

The advantage of using QFT to implement MCU relies on the efficient decomposition of controlled phase gates compared to $C-R_x(\pi/2^{m-1})$ used in the state-of-the-art linear-depth decomposition (LDD) of MCU \cite{Silva2022, Silva2023}. If we ignore noiseless $R_z$ gates and the two C-NOT gates used in both implementations, the decomposition of $C-R_z$ uses additional noisy $\sqrt{X}$ gates. However, both implementations use $C-U_m$ gates which are the most complex and hard to implement in a general case. A recent paper showed that the circuit can be simplified by omitting gates with $m>\lceil \log_2n\rceil$ \cite{Silva2023} similar to the approximate QFT approach. We should note that the QFT-based MCU circuit uses approximately the same number of non-elementary gates as the state-of-the-art one. However, the decomposition of $C-R_m$ in the NGS of superconducting hardware is twice as less complex as $C-R_x(\pi/2^{m-1})$ leading to approximately double the advantage of the QFT-based approach when implementing MCX \cite{Arsoski2024}. In the case of a general special unitary gate, the depth of the MCU circuit is predominantly determined by the complexity of the $C-U_m$ decomposition. Since our approach uses the same sequence of gates on the target qubit's wireline in the first ``triangular'' subcircuit, the difference in the LDD- and QFT-MCU depths is smaller than in the case of MCX-s. However, the number of elementary gates used in our approach is still approximately twice as small as in LDD.

\begin{figure}[ht]
  \centering
  \includegraphics[scale=0.75]{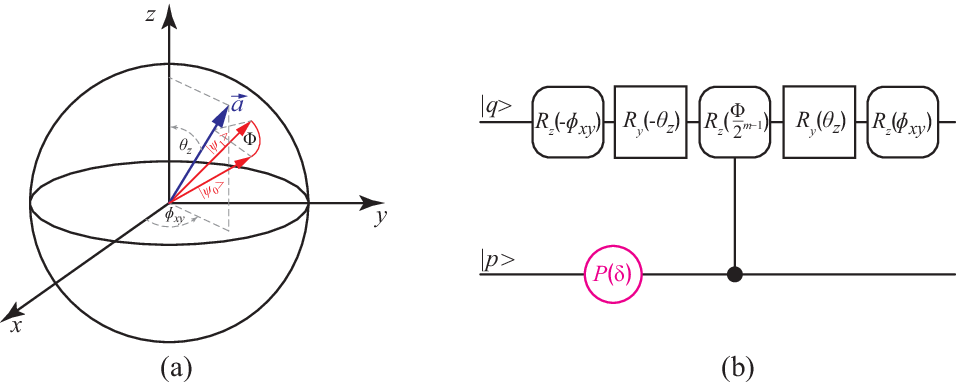}
  \caption{(a) Schematic representation of a general single-qubit state rotation by angle $\Phi$ about the axis defined by the vector $\vec{a}$ (displayed by a {\color{blue}blue arrow}). The angle between $\vec{a}$ and the $z$-axis is $\theta_z$, while the projection of $\vec{a}$ onto the $xy$-plane forms the angle $\phi_{xy}$ with the $x$-axis. The Bloch state vectors $\vert\psi_0\rangle$ and $\vert\psi_1\rangle=e^{-\i\frac{\Phi}{2}\vec{a}\vec{\sigma}}\vert\psi_0\rangle$, given by a {\color{red}red arrows}, denote the initial and the final single-qubit state up to the phase $e^{\i\delta}$, respectively. The trajectory of the state on the Bloch sphere, represented by a {\color{red}red arc}, is in the plane perpendicular to $\vec{a}$. (b) The decomposition of $C_p-U_{m,q}$.}
  \label{fig6}
\end{figure}

We introduce an alternative single-qubit unitary gate notation \cite{Nielsen2010} based on a general rotation about an arbitrary axis given by a unit vector $\vec{a}$ to optimize the QFT-MCU depth further. A single qubit unitary gate can be expressed as:
\begin{eqnarray}
	U&=&e^{\i\delta}R_{\vec{a}}(\Phi)=e^{\i\delta}e^{-\i\frac{\Phi}{2}\vec{a}\cdot\vec{\sigma}}=e^{\i\delta}R_z(\phi_{xy})R_y(\theta_z)R_z(\Phi)R_y(-\theta_z)R_z(\phi_{xy})\nonumber\\
	&=&e^{\i\delta}R_z(\phi_{xy})R_y(\theta_z)R_z(\Phi)(R_z(\phi_{xy})R_y(\theta_z))^\dagger,
\end{eqnarray}
where $\theta_z$ is the angle between $\vec{a}$ and the $z$-axis, and $\phi_{xy}$ is the angle between the projection of $\vec{a}$ in the $xy$-plane and the $x$-axis, as displayed in Fig.~\ref{fig6}(a). Thus, $R_z(-\phi_{xy})$ first rotates $\vec{a}$ into the $xz$-plane, and then $R_y(-\theta_z)$ executes rotation to align $\vec{a}$ with the $z$-axis. After performing the desired rotation about $\vec{a}$ by the angle $\Phi$ using $R_z(\Phi)$, the system is restored to its original orientation applying rotations $R_y(\theta_z)$ and $R_z(\phi_{xy})$, respectively. This decomposition is slightly more complex than the standard $U=e^{i\delta}R_z(\alpha)R_y(\theta)R_z(\beta)$. Using the general rotation formula from Ref.~\cite{Nielsen2010} it is straightforward to show that the alternative decomposition can be reduced to the standard one where $\sin{\frac{\theta}{2}}=\sin{\theta_z}\sin{\frac{\Phi}{2}}$, $\alpha=\alpha_z+\phi_{xy}$ and $\beta=\alpha_z-\phi_{xy}+\pi$ with $\tan{\alpha_z}=-\cot{\frac{\Phi}{2}} /\cos\phi_z$.

\begin{figure}[ht]
  \centering
  \includegraphics[scale=0.74]{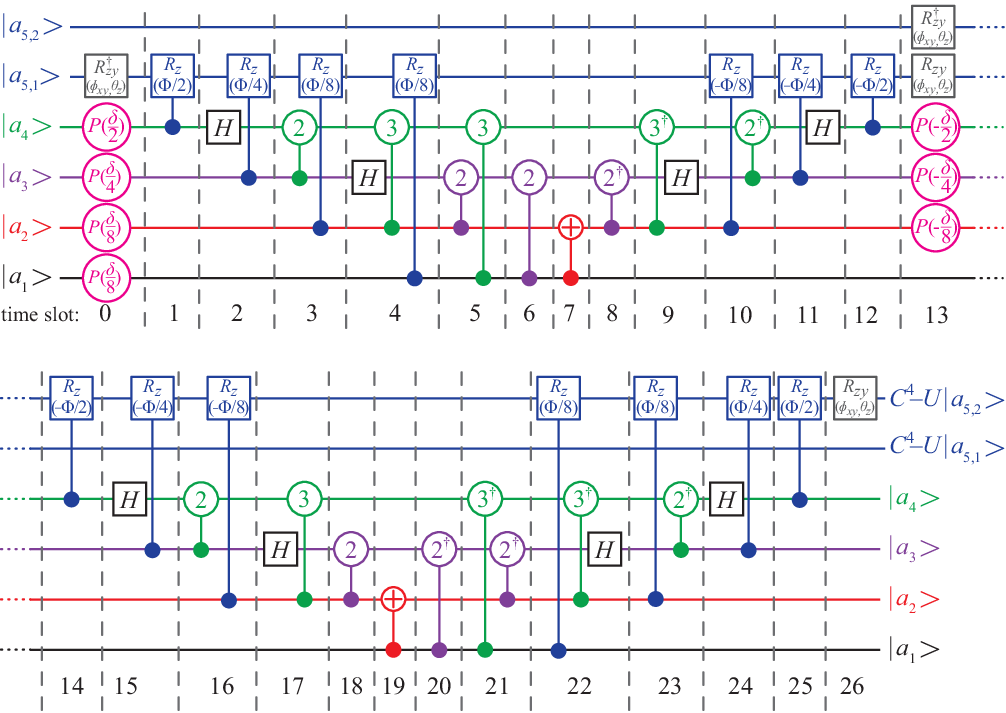}
  \caption{A schematic of the optimized circuit which is straightforwardly expanded to the multi-controlled multi-target unitary gate. We show the circuit with only two target qubits displayed in {\color{ultramarine}blue}. Instead of applying the sequence of $C-R_z(\Phi/2^{m-1})$ operations on two wirelines using only {\color{RoyalBlue}``$+1$''} subcircuit, we employed ``uncomputational'' {\color{red}``$-1$''} subcircuit  to implement $C-U$ to the second wireline.}
  \label{fig7}
\end{figure}

This alternative notation provides us with a rather simple symmetric form of $U^k$:
\begin{eqnarray}
	U^k&=&e^{\i k\delta}R^k_{\vec{a}}(\Phi)=e^{\i k\delta}e^{-\i\frac{\Phi}{2}k\vec{a}\cdot\vec{\sigma}}\nonumber\\
	&=&e^{\i k\delta}R_z(\phi_{xy})R_y(\theta_z)R_z(k\Phi)(R_z(\phi_{xy})R_y(\theta_z))^\dagger. \label{powerU}
\end{eqnarray}
To implement $C-U^k$ we use $C-R_z(k\Phi)$, as displayed in Fig.~\ref{fig6}(b). Although this decomposition is not less complex than the standard one, it will significantly simplify QFT-MCU. To implement our circuit,  we have to find
\begin{equation}
	U_m=U^{1/2^{m-1}}=e^{\i \delta/2^{m-1}}R_z(\phi_{xy})R_y(\theta_z)R_z(\Phi/2^{m-1})(R_z(\phi_{xy})R_y(\theta_z))^\dagger,
\end{equation}
where we used $k=1/2^{m-1}$ in Eq.~\ref{powerU}. Here, controlled $U_m$ gate is:
\begin{equation}
	C-U_m=(I_{2\times 2}\otimes R_{zy}(\phi_{xy},\theta_z))(C-R_z(\Phi/2^{m-1}))(I_{2\times 2}\otimes R_{zy}(\phi_{xy},\theta_z))^\dagger, \label{CUm}
\end{equation}
where $R_{zy}(\phi_{xy},\theta_z)=R_z(\phi_{xy})R_y(\theta_z)$. Using the above identity we may replace $C-{SU}_m$ gates with $C-R_z(\Phi/2^{m-1})$ in the {\color{RoyalBlue}``$+1$''} subcircuit, as displayed in Fig.~\ref{fig7}. One may note that $R_{zy}$ and $R_{zy}^\dagger$ between consecutive ${SU}_m$ gates acting on the target wireline in the {\color{RoyalBlue}``$+1$''} subcircuit cancel out. This significantly simplifies the circuit, since $C-R_z$ has approximately threefold simpler decomposition than a general $C-{SU}$. 

One may imply that this alternative notation ($U=U(\phi_{xy},\theta_z,\Phi,\delta)$) is very beneficial due to the symmetry argument and simplicity. It is straightforward to define many common gates, such as $X=U(0,\pi/2,\pi,\pi/2)$. Furthermore, some useful operations are simple to implement. For example, $C^{n-1}-U^\dagger(\phi_{xy},\theta_z,\Phi,\delta)=C^{n-1}-U(\phi_{xy},\theta_z,-\Phi,-\delta)$. 

If $\Phi$ is small, $R_z(\Phi/2^{m-1})$ approaches the identity matrix for a rather small value of $m$. In the place of these gates, QFT-MCX has $R_z(\pi/2^{m-1})$. When $\Phi$ is sufficiently smaller than $\pi$, MCU gates perform poorly compared to MCX. 

If we apply the sequence of gates acting on the target qubit to new target wirelines, we will implement the same unitary gate controllably. Thus, expanding QFT-MCU to the multi-controlled multi-target unitary gate (MCMTU) is straightforward. We can use {\color{RoyalBlue}``$+1$''} subcircuit to implement $C-U$ to as many qubits as needed, and {\color{red}``$-1$''$\:{\rm mod} \:2^{n-1}$} subcircuit is identical as in the MCU. In Fig.~\ref{fig7}, we have shown the modified 6-qubit MCU with four controls and two targets. We used {\color{RoyalBlue}``$+1$''} subcircuit to apply controlled unitary to the first target, and {\color{red}``$-1$''} to execute the same operation on the second target. This approach is beneficial if both target qubits are part of the qubits chain in the linear nearest-neighbor network. If we use only {\color{RoyalBlue}``$+1$''} subcircuit to implement $C-U$ on both targets, we would have to swap target qubits for applying each of $(2n-5)$ $C-R_z(\Phi/2^{m-1})$ operations. However, if we split execution between {\color{RoyalBlue}``$+1$''} and {\color{red}``$-1$''} subcircuits, we need to swap only twice, in slots 13 and 26 in Fig.~\ref{fig7}. If we apply a controlled operation to many target wirelines, the best option is to execute MCX on one clean ancilla qubit and then use it to implement $C-U$ on all targets.

\begin{figure}[ht]
  \centering
  \includegraphics[scale=0.7]{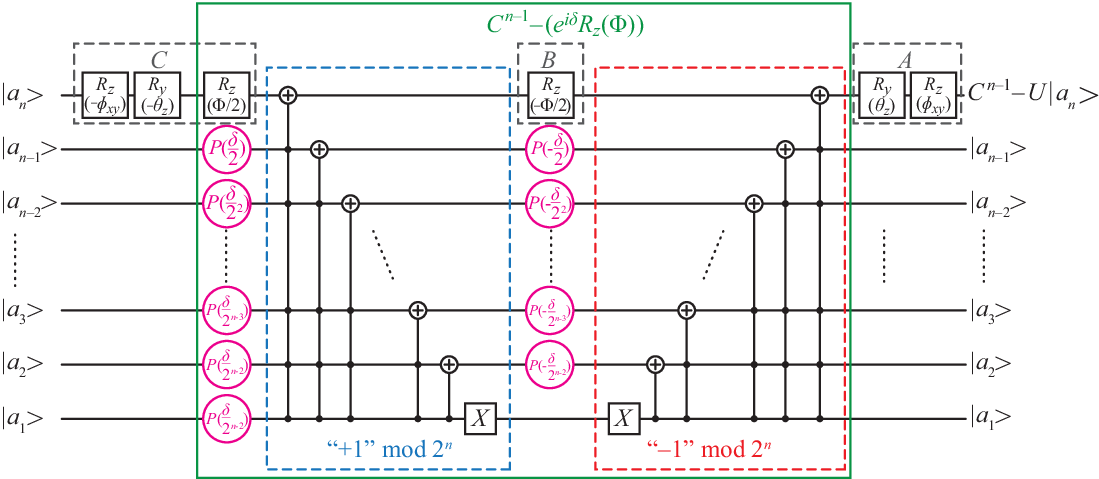}
  \caption{A schematic of $n$-qubit  MCU implementation based on the ZYZ-like decomposition and extended QFT-MCX. The phase gates that implement multi-controlled phase factor $e^{\i\delta}$ are represented by {\color{magenta}magenta circles}, while the {\color{ForestGreen}green rectangle} frames the part of the circuit which implements {\color{ForestGreen}$C^{n-1}-(e^{\i\delta}R_z(\Phi))$}. Groups of single-qubit gates acting on the target qubit, other than the ones belonging to MCX, are famed by {\color{gray}dashed gray rectangles} and denoted by $A$, $B$, and $C$, by analogy to the standard ZYZ decomposition.} 
  \label{fig8}
\end{figure}

The second implementation is also based on an alternative single-qubit notation. It is straightforward to conclude that a multi-controlled single-qubit unitary gate can be expressed as
\begin{eqnarray}
	C^{n-1}-U(\phi_{xy},\theta_z,\Phi,\delta) &=& [I_{(n-1)\times (n-1)}\otimes R_{zy}(\phi_{xy},\theta_z)]\nonumber\\
	&\cdot&[C^{n-1}-(e^{\i\delta}R_z(\Phi))]\nonumber\\
	&\cdot&[I_{(n-1)\times (n-1)}\otimes R_{zy}(\phi_{xy},\theta_z)]^\dagger,
\end{eqnarray}
where the decomposition of $C^{n-1}-R_z(\Phi)$ uses two MCX ($C^{n-1}-X$) gates
\begin{eqnarray}
	C^{n-1}-R_z(\Phi) = (C^{n-1}-X)(I_{(n-1)\times (n-1)}\otimes R_{z}(-\Phi/2))\nonumber\\
	\cdot(C^{n-1}-X)(I_{(n-1)\times (n-1)}\otimes R_{z}(\Phi/2)),
\end{eqnarray}
and the controlled phase factor can be implemented by adding appropriate phase gates to the controlled wirelines of MCXs, as explained in Section \ref{sec3}. To implement two MCX gates needed for the decomposition, we will use a single QFT-based MCX circuit with {\color{red}``$-1$''$\mod 2^n$} instead of $I_{2\times2}\otimes$({\color{red}``$-1$''$\mod 2^{n-1}$}), as shown in Fig.~\ref{fig8}. Thus, we used one MCX instead of the two needed in the standard approach. A simplified stair-wise MCX diagram used in MCU is displayed in Fig.~\ref{fig8}. A detailed representation of QFT-MCX circuits in FC and LNN architectures can be found in Ref.~\cite{Arsoski2024}. Here, $A=R_{zy}(\phi_{xy},\theta_z)=R_z(\phi_{xy})R_y(\theta_z)$, $B=R_z(-\Phi/2)$, and $C=(AB)^\dagger$. In the standard ZYZ decomposition, for a single-qubit unitary matrix expressed as $U=e^{i\delta}R_z(\alpha)R_y(\theta)R_z(\beta)$, there is a set of matrices $A=R_z(\alpha)R_y(\theta/2)$, $B=R_y(-\theta/2)R_z(-(\alpha+\beta)/2)$, and $C=R_z((\beta-\alpha)/2)$ such that $ABC=I_{2\times 2}$ and $U=e^{i\delta}AXBXC$ \cite{Barenco1995}. Thus, we use an alternative form of the ZYZ decomposition that is as complex as the standard one when implemented using the QFT-based approach. Therefore, an advantage of the ZYZ QFT-MCU implementation is not based on an alternative notation but on the fact that we use one QFT-MCX to implement MCU instead of the two MCXs used in the default approach. The complexity of the circuit is approximately equal to the QFT-MCX, where we use only additional $A$, $B$, and $C$ single-qubit gates. In general multi-controlled single-qubit $U(2)$ gate implementation, we have additional phase gates (shown in {\color{magenta}magenta} color in Fig.~\ref{fig8}), that implement $(n-1)$-qubit controlled phase gate. This doesn't affect the MCU's fidelity and time complexity, since {\color{magenta}$P$} gates are noiseless and can be executed parallel with $B$ and $C$. As in the case of QFT-MCU, it is straightforward to expand this circuit to MTMCU by applying the same sequence of gates on a new target wireline as on the $n$th.

The QFT-based subcircuits ({\color{RoyalBlue}``$+1$''} and {\color{red}``$-1$''}) implement two MCX gates and are executed in $(8n-12)$ time slots in the FC architecture. There are $4(n-2)$ Hadamard gates, $2n(n-2)$ controlled phases, and two $C-X$ gates. AQFT reduces the number of $C-R_m$ gates to  $2([\log_2 n]-1)(2n-1-[\log_2n])$. At most three time slots are required for executing $A$, $B$, and $C$. The LNN implementation uses an additional $(8n-16)$ time slots with $2(n-1)^2$ SWAP gates. Due to canceling out of two $C-X$ gates between SWAP and $C-R_z$ the increase in the number of elementary gates in LNN will be lower than it seems.

After reviewing the complexity metrics, we can ask ourselves if the second implementation is useful. However, using Lemma 7.9 and referring to Lemma 5.4 of Ref.~\cite{Barenco1995} implies that multi-controlled special unitary (MC-SU) gates can be efficiently simulated by splitting control qubits into two groups, as shown in Fig.~\ref{fig11}(a) for the SU $R_z$ gate used in the generalized notation. This approach is employed to linearise the number of C-NOT gates used in different MC-SU circuits \cite{Vale2023, Iten2016, He2017}. By roughly dividing the control register in half and utilizing dirty ancilla qubits in each group, the number of C-NOT gates in FC implementation is reduced to $(28n-88)$ for $n$ even and $(28n-92)$ for $n$ odd, provided the number of qubits is $n\geq 8$.\cite{Iten2016} Using additional optimization in Ref.~\cite{Vale2023} this number is reduced to $(20n-42)$ for $n$ even and $(20n-38)$ for $n$ odd, provided the number of qubits is $n\geq 5$.

\begin{figure}[ht]
  \centering
  \includegraphics[scale=0.8]{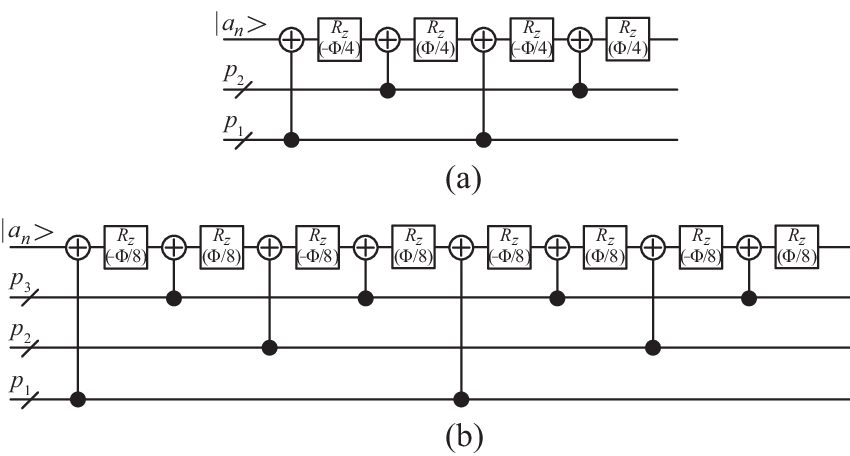}
  \caption{The decomposition of a multi-controlled $R_z$ gate, which is applied to the target in our approach that uses the alternative notation, when controls are divided into (a) two and (b) three groups.} 
  \label{fig9}
\end{figure}

Applying ABC decomposition recursively, it is straightforward to show that control qubits can be divided into more than two clusters, as shown in Fig.\ref{fig11}(b) for splitting controls into three groups. Using a deductive analysis, we found the simple binary rule to implement this circuit without an actual recursive decomposition. Providing that controls are divided into $m$ groups, there will be $2^{m}$ $R_z(\pm\Phi/2^{m})$ gates separated by MCX gates controlled by a certain group. The sequence starts with $R_z$ gate and ends with MCX, or vice versa. The nearest $R_z$s have the opposite sign of argument $(\Phi/2^{m})$. Labeling controlled groups by $p_{i}$, with $i$ ranging from 1 to $m$, and the position of MCX action on the target wireline from 1 to $2^{m}$, the first MCX should be applied to the first and $2^{m-1}$th position. Since one extended MCX implements two multi-controlled NOTs, we use only one MCX for this operation. Considering that {\color{red}``$-1$''} is inverse to {\color{RoyalBlue}``$+1$''}, the action of phase-adding gates cancels out and we use only relative MCX in implementation, as explained in section \ref{sec3}. The MCX gates controlled by the second cluster are applied periodically to positions labeled by $k\cdot2^{m-2}$th (where $k=\{1,...,2^2\}$) except the two positions ``occupied'' by MCX gate controlled by the first group. Therefore, the QFT-MCX circuit controlled by the second group is also used once to implement the two multi-controlled X on the $2^{m-2}$th and $3\cdot2^{m-2}$th position (that is, $2^2-2^1=2$). The third group is applied to the positions labeled by an integer multiple of $2^{m-3}$, excluding the one already occupied by lower indexed groups (namely, $2^3-2^2=4$ points and twice less QFT-MCX circuits, e.g. 2). Finally, the last group is applied to the remaining positions, which are $2^{m-1}$ locations labeled by an odd number. The MCXs controlled by the last cluster will be applied the most frequently and should contain the smallest number of controls, preferably one wireline when we use C-NOTs instead of MCX. 

\begin{figure}[ht]
  \centering
  \includegraphics[scale=0.142]{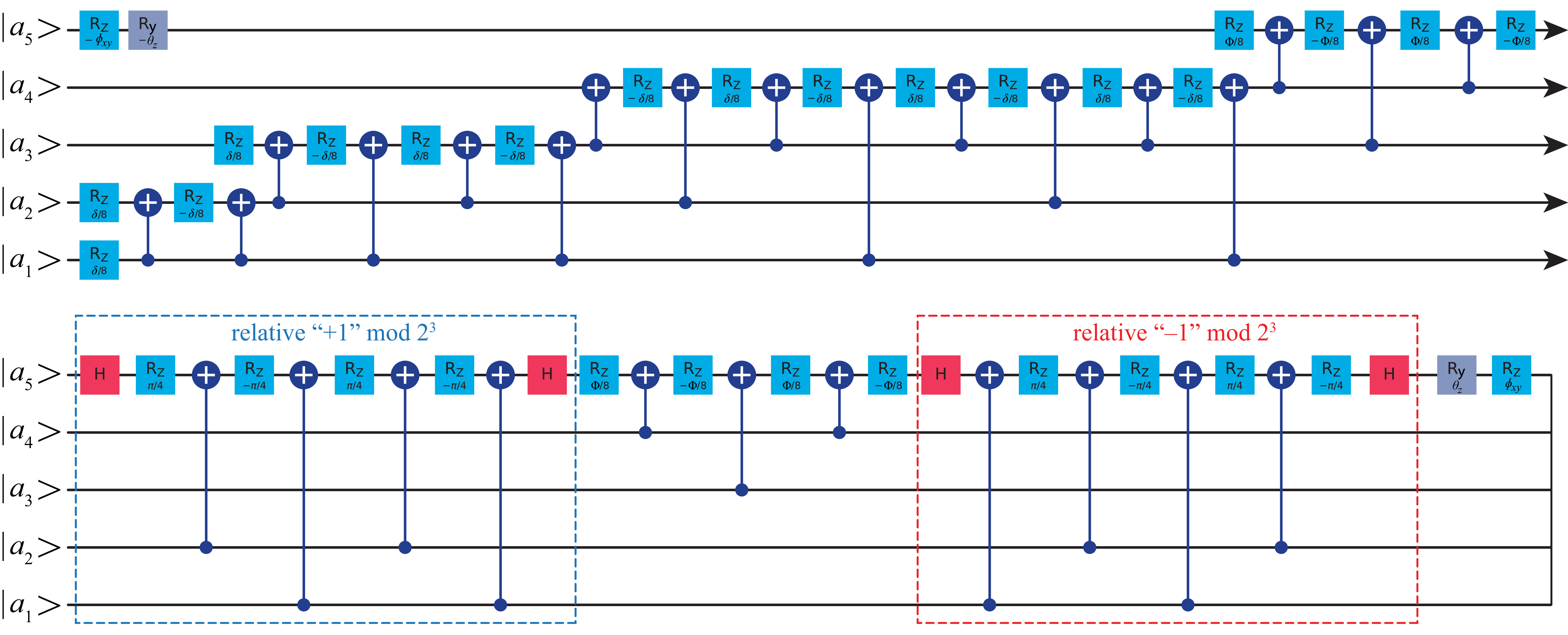}
  \caption{Optimal 5-qubit MCU based on the deep decomposition.}
  \label{fig10}
\end{figure}

We will use more than two groups if the price in the circuit depth and the number of C-NOT gates due to removing a single wireline from the larger of the first two groups is higher than adding one more cluster containing a single wireline. Moreover, the price of adding another wireline to the third cluster is higher than forming the fourth cluster containing a single wireline, and so on. Thus, the first two clusters must contain approximately equal to the number of control wirelines while others are comprised of a single wireline. From Fig.~\ref{fig8} it is apparent that a $R_z$ gate controlled by the lower qubits is implemented on each control wireline, except the first one. For example, the circuit acting on the $k$th wireline is $C^{k-1}-R_z(\delta/2^{n-k-1})$. Therefore, we can apply decomposition to phase circuits too. Thus, we can use this decomposition to optimize QFT-MCU. If circuits acting on the target and all control wire lines are decomposed as described, we will call this approach ``deep decomposition'' (DD). In Fig.~\ref{fig10}, we show the five-qubit DD-MCU. Rather straightforward but a more complex circuit for $n>4$ is when using $(n-1)$ groups each containing a single control when we apply by $2^{n-1}$ $R_z$ and $C-X$ gates of each on the target wireline. 

Building DD-MCU with more than four qubits requires MCX gates in the first two groups. These gates are comprised of {\color{RoyalBlue} ``$+1$''} and {\color{red} ``$-1$''} subcircuits, each executing the single MCX operation, as shown in Fig.~\ref{fig8}. If we use relative phase MCX, the gates in the second part of a circuit have to be inverted with respect to the ones in the first part. Then, the phase $-i$ in the first half of MCX cancels out with the $+i$ in the second half. We demonstrate this case in Fig.~\ref{fig10}, where we used {\color{RoyalBlue} relative ``$+1$''$\:{\rm mod}\: 2^3$} (the relative Toffoli) and {\color{red} relative ``$-1$''$\:{\rm mod}\: 2^3$} (the reversed relative Toffoli) circuits framed by dashed lines. Furthermore, each QFT-MCX gate applied to a wireline reduces the number of $z$-rotation operators needed by twice. Therefore, we double arguments in $R_z$ gates for each QFT-MCX employed for two multi-controlled X operations. This consideration is crucial when implementing the phase factor to the control wirelines.

The depth of this circuit is predominantly determined by the number of gates acting on the target wireline. The most complex circuit for $n>4$ is when having $(n-1)$ groups each containing a single control when we apply by $2^{n-1}$ $R_z$ and $C-X$ gates each to the target wireline.  In the next chapter, we will analyze circuit depth and the number of elementary gates used in various implementations. We will show that deep decomposition is useful. Although the increase in the number of C-NOT gates with the number of qubits is not linear in DD-circuits, up to the certain number of qubits used in MCU, the total number of C-NOTs is smaller than that in QFT-based circuits with a quadratic increase in the number of C-NOT gates.

\section{Analytical and numerical analysis}\label{sec5}

To get theoretical bounds for complexities in a genuine quantum computation, we adopt the native gate set $\{C-X, R_z, ID, SX=\sqrt{X}, X\}$, which is one of a few sets used by superconducting hardware. One may show that $R_m=\exp(i\frac{\pi}{2^m})\cdot R_z(\frac{\pi}{2^{m-1}})$, $H=\exp(i\frac{\pi}{4})\cdot R_z(\frac{\pi}{2})\sqrt{X} R_z(\frac{\pi}{2})$, $R_y=\sqrt{Z}HR_z H\sqrt{Z}^\dagger$, and $SWAP_{12}=(C_1-X_2)(C_2-X_1)(C_1-X_2)$\cite{Nielsen2010}. Hadamard and SWAP use three native gates and three elementary time intervals to execute, while $R_y$ and $C-R_m$ use five. Shifting and merging $R_z$ gates the depths of $C-R_z$ and $C-R_m$ gates are effectively reduced to three. Some of the $R_z$ gates can be executed simultaneously, as elaborated in Ref.~\cite{Arsoski2024}. Merging consecutive $R_z$ gates, $A$ and $C$ comprise 5 elementary gates, and $B$ uses one $R_z$ gate. There are $SWAP\cdot C-X$ and $SWAP\cdot C-R_z$ gate sequences in the LNN implementation, where two $C-X$ gates annihilate in each. In the most general case, the standard decomposition of a controlled single-qubit unitary gate in the NGS uses 14 elementary gates (two $C-X$, four $\sqrt{X}$, and the rest are $R_z$) that execute in 13 elementary time intervals. One should note that the decomposition of some $C-U$ gates (excluding $C-X$ or $C-Z$, since comprised in the basis gate set of superconducting quantum devices) is less complex, where $C-R_z$ and $C-P$ are some of the simplest. Merging $R_z$'s of neighboring $U_m$ gates, the depth and the number of elementary gates effectively reduce by one (except for the first or the last gate on the target wireline). However, using an alternative single-qubit representation we derived a method to substitute $C-U_m$ gates with $C-R_z(\Phi/2^{m-1})$ significantly reducing our circuit depth and the number of the elementary gates used.

Experimental analysis of LDD-MCX circuits demonstrated that omission of certain gates may result in a higher precision \cite{Silva2023}. However, a clear criterion for approximation has not been elaborated. In QFT-MCU we use $R_z(\pi/2^{m-1})$ and $R_z(\Phi/2^{m-1})$, where the former are a part of the QFT circuit while later are gates acting on the target qubit. To obtain a consistent approximation of the QFT-MCU circuit, we should keep $R_z(\varphi)$ gates with the argument $\varphi\geq \pi/2^{\lceil\log_2n\rceil-1}=\varphi_{\min}$. Therefore, the approximation would not be applicable if $\Phi<\varphi_{\min}$. However, we may choose the optimal $\varphi_{\min}$ for our circuit.

We will estimate the lower and upper bounds for the circuit depth and number of elementary gates for a multi-controlled $U(2)$ gate. All bounds, expressed as a function of the number of qubits used, are derived employing the selected NGS. We will use all the optimizations above thus obtaining the minimal values for complexity bounds. If the transpiling software does not find the optimal conditions, the circuit complexities may exceed the calculated upper bounds. 

By counting in the NGS, FC-MCU based on MCX modification has the depth $(24n-53)$ for $n>3$. The circuit uses $4(n-2)$ $\sqrt{X}$ or $\sqrt{X}^\dagger$, and $4(n-1)(n-2)$ $C-X$ gates. In the fully optimized circuit, the number of $R_z$ gates is at least $(2n^2-17)$. An approximation of the circuit depends on $\Phi$ and may additionally reduce the number of C-NOTs and $R_z$ gates. Due to the SWAP gates used in the LNN implementation, the depth of the circuit increases by at least $(8n-18)$ for $n>3$. We do not need to swap qubits for $n=3$. Also, the number of $C-X$ gates enlarges by at least $(2n^2-6n+6)$. For a general $C-U(2)$ circuit, we use $(2n-3)$ $R_z$ gates to implement the controlled-phase circuit.

FC-MCU based on the ZYZ-like decomposition has the depth $(24n-31)$. It uses $2(n^2+2n-6)$ $R_z$, $4(n-1)$ $\sqrt{X}$ or $\sqrt{X}^\dagger$, and $2(2n^2-4n+1)$ $C-X$ gates. An approximation of the circuit is straightforward, as in the case of standard QFT. It will reduce the number of $R_z$ and $C-X$ gates to $2(2(2n-3)+([\log_2n]-1)(2n-1-[\log_2n]))$ and $2(1+2([\log_2n]-1)(2n-1-[\log_2n]))$, respectively. Additional SWAP gates used in the LNN architecture increase circuit depth and the number of $C-X$ gates by at most $(8n-16)$ and $2(n-1)^2$ for $n>3$, respectively. This circuit also uses $(2n-3)$ $R_z$ gates to implement $e^{i\delta}$.

    \begin{table}[h]
	\caption{Qubits clustering and the number of C-NOTs in $n$-qubit DD-MCUs.}\label{tab1}%
        \begin{tabular}{@{}lllll@{}}
        \toprule
        $n$ & The number &  Controls per &  The number of&  The number of\\ 
         qubits &  of clusters & cluster & C-NOTs in MC-SU(2)\footnotemark[1] & C-NOTs in MC-U(2)\footnotemark[2] \\
        \midrule
        3 & 2 & [1, 1] & 4 & 6 \\
	4 & 3 & [1, 1, 1] & 8 & 14 \\
	5 & 3 & [2, 1, 1] & 14 & 28 \\
	6 & 3 & [2, 2, 1] & 20 & 48 \\
	7 & 3 & [3, 2, 1] & 28 & 76 \\
	8 & 3 & [3, 3, 1] & 36 & 112 \\
	9 & 4 & [3, 3, 1, 1] & 44 & 156 \\
	10 & 4 & [4, 3, 1, 1] & 60 & 216 \\
	11 & 4 & [4, 4, 1, 1] & 76 & 292 \\
	12 & 5 & [4, 4, 1, 1, 1] & 92 & 384 \\
	13 & 5 & [5, 4, 1, 1, 1] & 124 & 508 \\
	14 & 5 & [5, 5, 1, 1, 1] & 156 & 664 \\
	15 & 6 & [5, 5, 1, 1, 1, 1] & 188 & 852 \\
        \botrule
    \end{tabular}
\footnotetext{Comparison to other methods.}
\footnotetext[1]{For $n\leq15$ is lower than $(16n-40)$ in the most optimal SU(2) implementation.\cite{Silva2023}}
\footnotetext[2]{For $n\leq12$ is lower than $4(n-1)(n-3)$ in QFT-MCU.}
\end{table}

As shown in the previous chapter, the number of elementary gates in a DD-MCU is non-linear with the number of qubits. In Table \ref{tab1} we give the number of C-NOT gates used in DD-MC-SU(2) and DD-MC-U(2) circuits as a function of the number of qubits $n$. One may show that the number of C-NOTs used in DD-MCU is lower than that employed in QFT-MCU for $n\leq12$. Moreover, the most optimal MC-SU(2) circuit \cite{Silva2023} uses more C-NOTs than DD-MCU for $n\leq15$. We consider only C-NOT or $C-Z$ gates since the median error of these gates is approximately 20 times higher than $\sqrt{X}$ used in our circuits, while $R_z$ gates are error-free. Based on these facts and the current performance of NISQ devices, we infer that DD-MCU can have the highest impact on practical applications.

\begin{figure}[ht]
  \centering
  \includegraphics[scale=0.65]{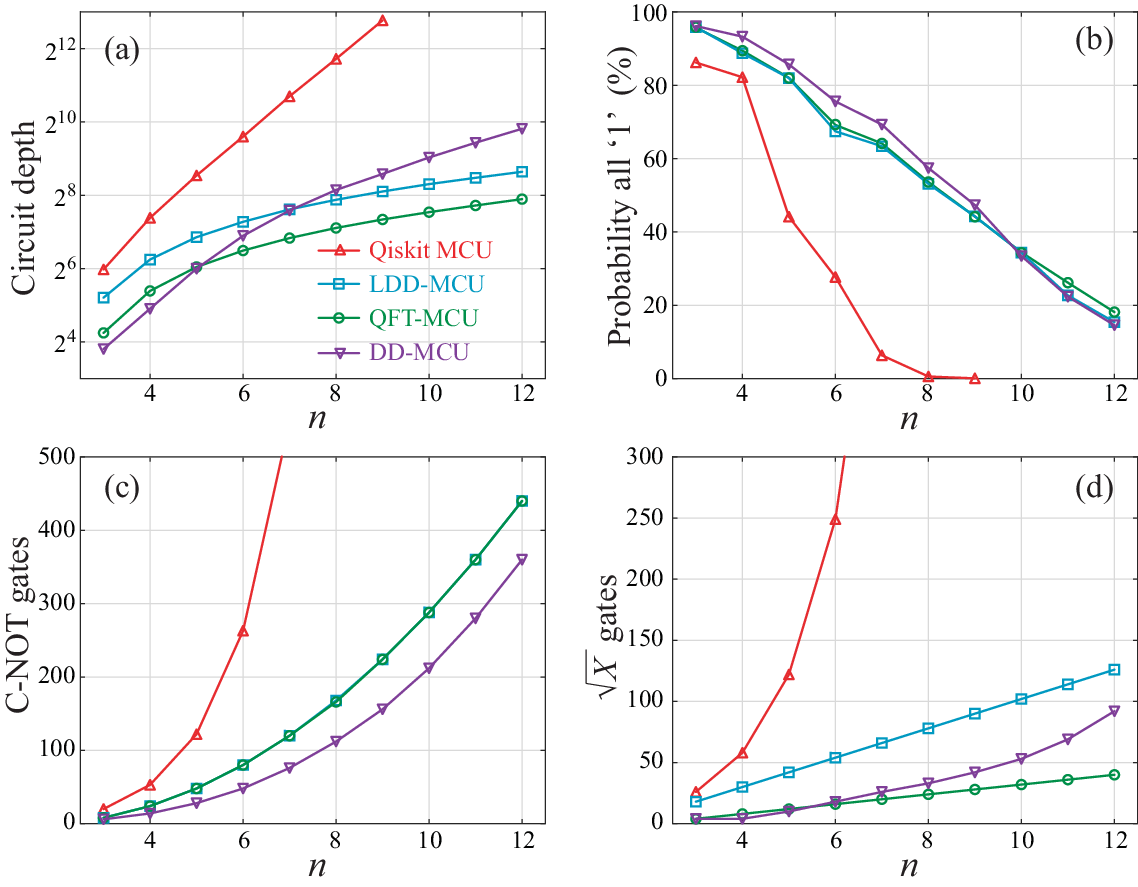}
  \caption{The dependence of (a) circuit depths, (b) probabilities for measuring  `1' at all outputs, (c) the number of C-NOT and (d) $\sqrt{X}$ gates on the number of qubits used in MCU in the {\color{red} default Qiskit} ({\color{red}$\bigtriangleup$}), {\color{brightcerulean}LDD-MCU} from Ref.~\cite{Silva2022} ({\color{brightcerulean}$\Box$}),   {\color{ForestGreen}QFT-MCU} ({\color{ForestGreen}$\bigcirc$}), and {\color{lavenderindigo}DD-MCU} ({\color{lavenderindigo}$\bigtriangledown$}). Due to the high complexity of the default Qiskit circuit, the software failed to emulate results for $n>9$.}
  \label{fig11}
\end{figure}

Actual complexities for execution on a quantum device are obtained by assembling analyzed circuits using the transpile function built-in Python package Qiskit v.1.3.2 \cite{Qiskit}. In doing so, we employed optimization level 3. We used the local simulator to transpile circuits for application on the latest version of a generic quantum device. The Qiskit Runtime local testing mode is engaged to emulate noisy MCU calculations. To find circuit reliability, we set all input qubits to $\vert 1\rangle$ and appended the target qubit with $U^\dagger$, as in Ref.~\cite{Silva2022}. We consider the calculation to be executed properly only if we measure `one' at all the outputs. Emulations were repeated $n_{shots}=10^5$ times on each circuit. We have compared our implementations to the standard Qiskit and the state-of-the-art LDD-MCU circuit. The unitary single-qubit gate used in all analyzed MCU circuits was randomly chosen from $U(2)$ gates. To get angles $\phi_{xy}$, $\theta_z$, $\delta$ and $\Phi$, we have generated random rational numbers and multiplied them by $\pi$. We constrained our choice to exclude trivial gates. In Figs.~\ref{fig11} (a) and (b), we show comparative results of variation in circuit depth and reliability with the number of qubits, respectively. Furthermore, we compare the number of C-NOT and $\sqrt{X}$ gates used in various implementations in Fig.~\ref{fig11} (c) and (d), respectively.

The default Qiskit implementation exhibits an exponential increase in the time complexity with the number of qubits, while circuit depths of either LDD or QFT-based MCUs are linear. LDD and QFT-MCU use $C-U_m$ gates, which have a complex decomposition that increases the circuit depth compared to MCX implementation. Using an alternative notation, we simplified $C-U_m$ to $C-R_z$ in our approach. Moreover, LDD uses $C-R_x$ gates that are more complex than $C-R_m$ comprising QFT-based implementation. Therefore, the circuit depths of LDD-MCU are twofold larger than QFT-MCU even though they have a similar construction. Although not linear, DD-MCU has the smallest depth up to a five-qubit circuit, as shown in Fig.~\ref{fig11}(a).

Since the default Qiskit MCU is the most complex, it exhibits the lowest fidelity of all other implementations (see Fig.~\ref{fig11}(b)). Moreover, there is an exponential increase in the number of C-NOT and $\sqrt{X}$ gates used, as displayed in Figs.\ref{fig11}(c) and (d), respectively. Detail analysis shows that the probability of measuring `1' at all outputs in the default Qiskit MCU is approximately $n_{shots}/2^n$ for $n\geq7$ implying a white noise-like output in the circuit. Up to $n=10$, DD-MCU has the highest fidelity of all. This is a consequence of using the smallest number of noisy C-NOT gates, as can be inferred from Fig.~\ref{fig11}(c). However, since the depth of the circuit increases non-linearly and significantly exceeds the value of the depth of the QFT and LDD circuits for $n>10$, it becomes less reliable than the later two implementations. QFT-MCU circuit exhibits a small advantage in fidelity over LDD-MCU. One may note that both implementations use an equal number of noisy C-NOT gates (see Fig.~\ref{fig11}(c)). However, LDD-MCU has a twofold higher depth and uses approximately three times more low-noisy $\sqrt{X}$ gates than our QFT-MCU, which becomes significant for $n>10$. Considering these facts, and following the discussion in section \ref{sec3}, one may wonder if LDD-MCU is somehow equivalent to our QFT-MCU. Straightforward methods to simplify LDD-MCU to our QFT-MCU are elaborated in Appendix \ref{sec7}. One should be noted that the results for LDD are in excellent agreement with the experimental ones (see the lower panel of Fig.~4 in Ref.~\cite{Silva2022}). Therefore, the emulation of noisy quantum calculations should be comparable to genuine quantum computations.

\begin{figure}[ht]
  \centering
  \includegraphics[scale=0.43]{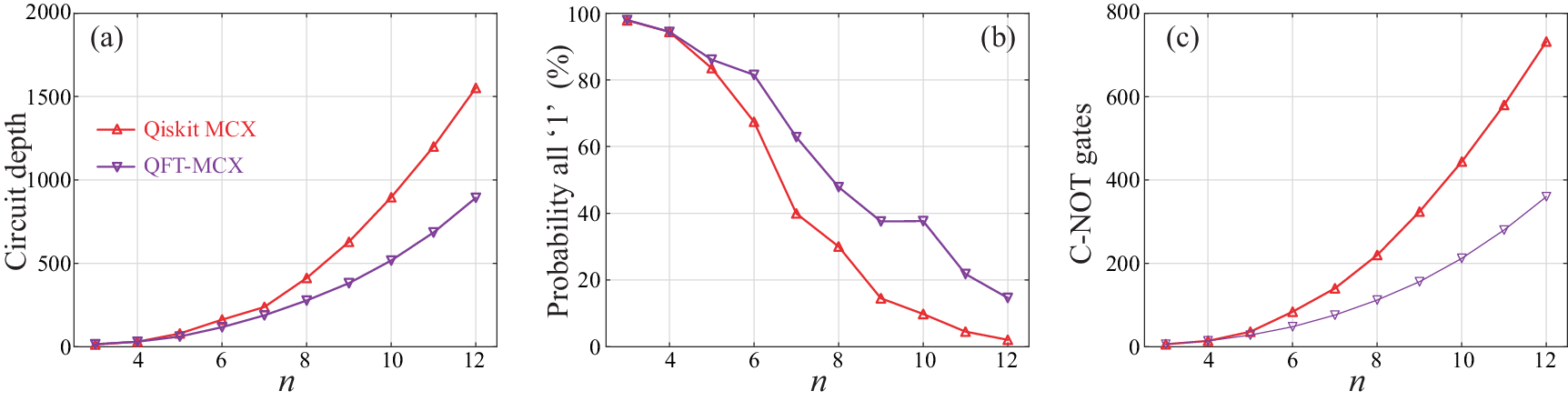}
  \caption{The dependence of (a) circuit depths, (b) probability for measuring  `1' at all outputs, and (c) the number of C-NOT gates in a circuit on the number of qubits used by MCU in the {\color{red} state-of-the-art Qiskit's MCX} ({\color{red}$\bigtriangleup$}) and {\color{lavenderindigo} DD-MCX} ({\color{lavenderindigo}$\bigtriangledown$}).}
  \label{fig12}
\end{figure}

Finally, we use our DD-MCU to implement a multi-controlled X gate and compare it to the Qiskit implementation. One should note that Qiskit's library contains custom-tailored low-depth MCX circuits using a small number of C-NOTs. Their schematics can be found in state-of-the-art references. In Fig.~\ref{fig12}, we compare circuit depths, fidelities, and the number of C-NOT gates used in our DD and Qiskit MCX implementation. For circuits with up to four qubits, both implementations have the same depth, fidelity, and number of C-NOT gates used. Starting from 5-qubit MCX, our deep decomposed circuit is less complex (see panel (a)), more reliable (panel (b)), and uses fewer C-NOT gates (panel (c)) than state-of-the-art Qiskit's MCX. 

Our library used to generate results presented in this work is publicly available on GitHub \cite{qccirc}. Given the high fidelities and low depths of our QFT-based circuits, we assert that they implement multi-controlled circuits in the qubit-based devices as well as a perspective higher-dimensional qudit-based hardware \cite{Zheng2022, Saha2022}.

\section{Conclusions}\label{sec6}

In this paper, we have presented a few new implementations of multi-controlled unitary (MCU) gates based on the modification of multi-controlled X (MCX) circuits that use the quantum Fourier transform (QFT). In the first implementation, the phase gates acting on the target qubit ($Z^{1/2^{m-1}}$) of QFT-MCX, are replaced with the adequate roots of the single-qubit unitary gate ($U^{1/2^{m-1}}$). The main drawback of this circuit is that it uses $U_m=U^{1/2^{m-1}}$ gates, which are relatively complex to implement. However, in our approach we used an alternative single-qubit gate notation to simplify $C-U_m$ gates to $C-R_z$, thereby making our circuit as simple as MCX. The second implementation is based on the ZYZ-like decomposition and uses an extended QFT-based MCX circuit to implement the two MCX gates needed for the decomposition. Further simplifications of these two are straightforward by approximating QFT or introducing auxiliary qubits. We used our second implementation recursively on the QFT-MCU with control qubits divided into clusters. It provides us with a deep decomposition (DD) of our QFT-MCU. We showed that DD-MCU uses the least number of noisy $C-X$ gates, which makes it the most useful for application in the current noisy quantum devices. Our implementations have lower time and space complexities compared to any existing MCU. Running noisy emulations on transpiled circuits demonstrated that our MCUs have a noticeable advantage in fidelity compared to state-of-the-art implementations.

We elaborated various techniques to optimize our circuits. Moreover, we demonstrated that the state-of-the-art linear-depth decomposition (LDD) MCU can be simplified to our QFT-MCU, thus significantly reducing the LDD's time and space complexities. Our alternative notation may facilitate the optimization and simplification of various complex quantum circuits.

\begin{appendices}
\setcounter{figure}{12}
\renewcommand{\thefigure}{\arabic{figure}}

\setcounter{equation}{0}

\section{Optimization of state-of-the-art LDD-MCU circuit}\label{sec7}

\begin{figure}[ht]
  \centering
  \includegraphics[scale=0.65]{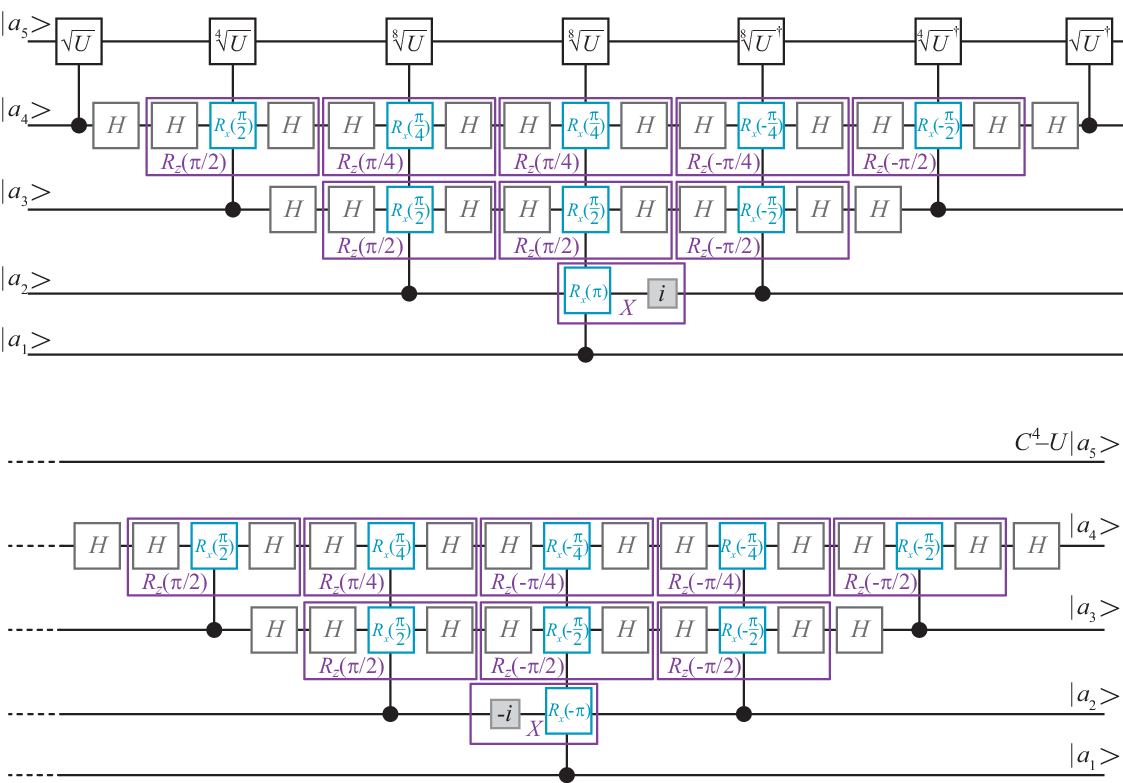}
  \caption{Simplification of {\color{brightcerulean}LDD-MCU} to our {\color{lavenderindigo} QFT-MCU}. Pairs of Hadamard gates (${\color{gray}HH}=I_{2\times 2}$) added to the {\color{brightcerulean}LDD-MCU} are shown by gray-outlined squares ({\color{gray}$\Box$}), where the two constant phase gates $\pm i I_{2\times 2}$ are given by small gray squares ({\color{gray}$\blacksquare$}).}
  \label{fig13}
\end{figure}

We found that if we insert two consecutive Hadamard gates between $R_x$ gates from References~\cite{Silva2022, Silva2023} (where $HH=I_{2\times 2}$), and using the equivalence $HR_x(\gamma)H=R_z(\gamma)$, all the $C-R_x(\gamma)$ gates will transform to $C-R_z(\gamma)$. On the second wireline $R_x(\pm \pi)=\mp i X$. We can add two constant phase gates, $+i I_{2\times 2}$ and $-i I_{2\times 2}$, between $R_x$ gates transforming $C-R_x(\pm\pi)$ to $C-X$ gates. By doing this, we transitioned from the $X$-basis used in LDD to the $Z$-basis utilized in our approach. When we perform all transformations above, each control wireline, except the first two, will have one $H$ gate `remaining' at the beginning and the end of the $C-R_z$ sequence. Thus, the LDD-MCU is reduced to our QFT-MCU from Fig.~\ref{fig5}, as demonstrated in Fig.~\ref{fig13}. This will decrease the number of elementary gates employed and the LDD circuit's depth by approximately twice. 

It is straightforward to show that equivalent simplification can be obtained using our alternative single-qubit notation applied to $C-U_m$ and $C-R_x$ gates. We imply that this method may effectively simplify many state-of-the-art circuits.

Finally, in Ref.~\cite{Silva2023} authors shown that using decomposition in Fig.~\ref{fig9}(a), Lemma 7.2 with Corollary 7.4 from Ref.~\cite{Barenco1995}, and Lemma 8 from Ref.~\cite{Iten2016}, multi-controlled SU(2) circuit can be implemented with at most $(16n-40)$ C-NOTs for $n\geq5$. To implement U(2) gates, a series of multi-controlled $R_z$ gates should be implemented on control wirelines. Thus, on the $(n-1)$th wireline we have to implement $C^{n-1}-R_z(\delta)$ gate, which implies use of $16(n-1)-40$ C-NOTs. On the $(n-2)$th wireline we implement $C^{n-2}-R_z(\delta/2)$ utilizing $16(n-2)-40$ C-NOTs. Finally, we implement the circuit with $(16\cdot4 - 40)$ C-NOTs on the first five qubits. The total number of C-NOT gates employed in MCU is:
\begin{eqnarray}
	&&(16n-40)+(16(n-1)-40)+(16(n-2)-40)+\cdots + (16\cdot4 - 40) \nonumber\\
	&=& 16(n+(n-1)+\cdots+5+4)-40(n-3)\nonumber\\
	&=&4(n-1)(n-3),
\end{eqnarray}
which is the number of C-NOTs used in the QFT-MCU.
\end{appendices}

\bigskip
\backmatter

\bmhead{Acknowledgements}

This work was financially supported by the Ministry of Science, Technological Development and Innovation of the Republic of Serbia under contract number: 451-03-65/2024-03/200103.

\bigskip


\begin{thebibliography}{99}
\ifx \bisbn   \undefined \def \bisbn  #1{ISBN #1}\fi
\ifx \binits  \undefined \def \binits#1{#1}\fi
\ifx \bauthor  \undefined \def \bauthor#1{#1}\fi
\ifx \batitle  \undefined \def \batitle#1{#1}\fi
\ifx \bjtitle  \undefined \def \bjtitle#1{#1}\fi
\ifx \bvolume  \undefined \def \bvolume#1{\textbf{#1}}\fi
\ifx \byear  \undefined \def \byear#1{#1}\fi
\ifx \bissue  \undefined \def \bissue#1{#1}\fi
\ifx \bfpage  \undefined \def \bfpage#1{#1}\fi
\ifx \blpage  \undefined \def \blpage #1{#1}\fi
\ifx \burl  \undefined \def \burl#1{\textsf{#1}}\fi
\ifx \doiurl  \undefined \def \doiurl#1{\url{https://doi.org/#1}}\fi
\ifx \betal  \undefined \def \betal{\textit{et al.}}\fi
\ifx \binstitute  \undefined \def \binstitute#1{#1}\fi
\ifx \binstitutionaled  \undefined \def \binstitutionaled#1{#1}\fi
\ifx \bctitle  \undefined \def \bctitle#1{#1}\fi
\ifx \beditor  \undefined \def \beditor#1{#1}\fi
\ifx \bpublisher  \undefined \def \bpublisher#1{#1}\fi
\ifx \bbtitle  \undefined \def \bbtitle#1{#1}\fi
\ifx \bedition  \undefined \def \bedition#1{#1}\fi
\ifx \bseriesno  \undefined \def \bseriesno#1{#1}\fi
\ifx \blocation  \undefined \def \blocation#1{#1}\fi
\ifx \bsertitle  \undefined \def \bsertitle#1{#1}\fi
\ifx \bsnm \undefined \def \bsnm#1{#1}\fi
\ifx \bsuffix \undefined \def \bsuffix#1{#1}\fi
\ifx \bparticle \undefined \def \bparticle#1{#1}\fi
\ifx \barticle \undefined \def \barticle#1{#1}\fi
\bibcommenthead
\ifx \bconfdate \undefined \def \bconfdate #1{#1}\fi
\ifx \botherref \undefined \def \botherref #1{#1}\fi
\ifx \url \undefined \def \url#1{\textsf{#1}}\fi
\ifx \bchapter \undefined \def \bchapter#1{#1}\fi
\ifx \bbook \undefined \def \bbook#1{#1}\fi
\ifx \bcomment \undefined \def \bcomment#1{#1}\fi
\ifx \oauthor \undefined \def \oauthor#1{#1}\fi
\ifx \citeauthoryear \undefined \def \citeauthoryear#1{#1}\fi
\ifx \endbibitem  \undefined \def \endbibitem {}\fi
\ifx \bconflocation  \undefined \def \bconflocation#1{#1}\fi
\ifx \arxivurl  \undefined \def \arxivurl#1{\textsf{#1}}\fi
\csname PreBibitemsHook\endcsname

\bibitem[\protect\citeauthoryear{Preskill}{2018}]{Preskill2018}
\begin{barticle}
\bauthor{\bsnm{Preskill}, \binits{J.}}:
\batitle{Quantum Computing in the NISQ era and beyond}.
\bjtitle{Quantum}
\bvolume{2},
\bfpage{79}
(\byear{2018}).
\doiurl{10.22331/q-2018-08-06-79}
\end{barticle}
\endbibitem

\bibitem[\protect\citeauthoryear{Kim et~al.}{2023}]{Kim2023}
\begin{barticle}
\bauthor{\bsnm{Kim}, \binits{Y.}},
\bauthor{\bsnm{Eddins}, \binits{A.}},
\bauthor{\bsnm{Anand}, \binits{S.}},
\bauthor{\bsnm{Wei}, \binits{K.X.}},
\bauthor{\bsnm{van den Berg}, \binits{E.}},
\bauthor{\bsnm{Rosenblatt}, \binits{S.}},
\bauthor{\bsnm{Nayfeh}, \binits{H.}},
\bauthor{\bsnm{Wu}, \binits{Y.}},
\bauthor{\bsnm{Zaletel}, \binits{M.}},
\bauthor{\bsnm{Temme}, \binits{K.}},
\bauthor{\bsnm{Kandala}, \binits{A.}}:
\batitle{Evidence for the utility of quantum computing before fault tolerance}.
\bjtitle{Nature}
\bvolume{618},
\bfpage{500}--\blpage{505}
(\byear{2023}).
\doiurl{10.1038/s41586-023-06096-3}
\end{barticle}
\endbibitem

\bibitem[\protect\citeauthoryear{Plesch}{2011}]{Plesch2011}
\begin{barticle}
\bauthor{\bsnm{Plesch}, \binits{M.}},
\bauthor{\bsnm{Brukner}, \binits{\v{C}.}}:
\batitle{Quantum-state preparation with universal gate decompositions}.
\bjtitle{Phys. Rev. A}
\bvolume{83}(\bissue{3}),
\bfpage{032302 }
(\byear{2011}).
\doiurl{10.1103/PhysRevA.83.032302}
\end{barticle}
\endbibitem

\bibitem[\protect\citeauthoryear{Zhang}{2021}]{Zhang2021}
\begin{barticle}
\bauthor{\bsnm{Zhang}, \binits{X.-M.}},
\bauthor{\bsnm{Yung}, \binits{M.-H.}},
\bauthor{\bsnm{Yuan}, \binits{X.}}:
\batitle{Low-depth quantum state preparation}.
\bjtitle{Phys. Rev. Res.}
\bvolume{3}(\bissue{4}),
\bfpage{043200 }
(\byear{2021}).
\doiurl{10.1103/PhysRevResearch.3.043200}
\end{barticle}
\endbibitem

\bibitem[\protect\citeauthoryear{Araujo}{2021}]{Araujo2021}
\begin{barticle}
\bauthor{\bsnm{Araujo}, \binits{I.F.}},
\bauthor{\bsnm{Park}, \binits{D.K.}},
\bauthor{\bsnm{Petruccione}, \binits{F.}},
\bauthor{\bsnm{da Silva}, \binits{A.J.}}:
\batitle{A divide-and-conquer algorithm for quantum state preparation}.
\bjtitle{Sci. Rep.}
\bvolume{11}(\bissue{1}),
\bfpage{6329 }
(\byear{2021}).
\doiurl{10.1038/s41598-021-85474-1}
\end{barticle}
\endbibitem

\bibitem[\protect\citeauthoryear{Barenco et.~al}{1995}]{Barenco1995}
\begin{barticle}
\bauthor{\bsnm{Barenco}, \binits{A.}},
\bauthor{\bsnm{Bennett}, \binits{C.H.}},
\bauthor{\bsnm{Cleve}, \binits{R.}},
\bauthor{\bsnm{DiVincenzo}, \binits{D.P.}},
\bauthor{\bsnm{Shor}, \binits{P.}},
\bauthor{\bsnm{Sleator}, \binits{T.}},
\bauthor{\bsnm{Smolin}, \binits{J.A.}},
\bauthor{\bsnm{Weinfurter}, \binits{H.}}:
\batitle{Elementary gates for quantum computation}.
\bjtitle{Phys. Rev. A}
\bvolume{52}(\bissue{5}),
\bfpage{3457}--\blpage{3467}
(\byear{1995}).
\doiurl{10.1103/PhysRevA.52.3457}
\end{barticle}
\endbibitem

\bibitem[\protect\citeauthoryear{Nielsen and Chuang}{2010}]{Nielsen2010}
\begin{bbook}
\bauthor{\bsnm{Nielsen}, \binits{M.C.}},
\bauthor{\bsnm{Chuang}, \binits{I.L.}}:
\bbtitle{Quantum Computation and Quantum Information}.
\bpublisher{Cambridge University Press},
\blocation{New {Y}ork}
(\byear{2010}).
\bisbn{978-1-107-00217-3}
\end{bbook}
\endbibitem


\bibitem[\protect\citeauthoryear{Shende}{2006}]{Shende2006}
\begin{barticle}
\bauthor{\bsnm{Shende}, \binits{V.V.}},
\bauthor{\bsnm{Bullock}, \binits{S.S.}},
\bauthor{\bsnm{Markov}, \binits{I.L.}}:
\batitle{Synthesis of quantum-logic circuits}.
\bjtitle{IEEE Transactions on CAD}
\bvolume{25}(\bissue{6}),
\bfpage{1000}--\blpage{1010}
(\byear{2006}).
\doiurl{10.1109/TCAD.2005.855930}
\end{barticle}
\endbibitem

\bibitem[\protect\citeauthoryear{Malvetti}{2021}]{Malvetti2021}
\begin{barticle}
\bauthor{\bsnm{Malvetti}, \binits{E.}},
\bauthor{\bsnm{Iten}, \binits{R.}},
\bauthor{\bsnm{Colbeck}, \binits{R.}}:
\batitle{Quantum circuits for sparse isometries}.
\bjtitle{Quantum}
\bvolume{5},
\bfpage{412 }
(\byear{2021}).
\doiurl{10.22331/q-2021-03-15-412}
\end{barticle}
\endbibitem

\bibitem[\protect\citeauthoryear{Bae et.~al}{2020}]{Bae2020}
\begin{barticle}
\bauthor{\bsnm{Bae}, \binits{J.-H.}},
\bauthor{\bsnm{Alsing}, \binits{P.M.}},
\bauthor{\bsnm{Ahn}, \binits{D.}},
\bauthor{\bsnm{Miller}, \binits{W.A.}}:
\batitle{Quantum circuit optimization using quantum Karnaugh map}.
\bjtitle{Sci. Rep.}
\bvolume{10}(\bissue{1}),
\bfpage{15651 }
(\byear{2020}).
\doiurl{10.1038/s41598-020-72469-7}
\end{barticle}
\endbibitem

\bibitem[\protect\citeauthoryear{Brugiere et.~al}{2021}]{Brugiere2021}
\begin{barticle}
\bauthor{\bsnm{de Brugi\`ere}, \binits{T.G.}},
\bauthor{\bsnm{Baboulin}, \binits{M.}},
\bauthor{\bsnm{Valiron}, \binits{B.}},
\bauthor{\bsnm{Martiel}, \binits{S.}},
\bauthor{\bsnm{Allouche}, \binits{C.}}:
\batitle{Reducing the depth of linear reversible quantum circuits}.
\bjtitle{IEEE Trans. Quantum Eng.}
\bvolume{2},
\bfpage{3102422 }
(\byear{2021}).
\doiurl{10.1109/TQE.2021.3091648}
\end{barticle}
\endbibitem

\bibitem[\protect\citeauthoryear{Cuomo et.~al}{2023}]{Cuomo2023}
\begin{barticle}
\bauthor{\bsnm{Cuomo}, \binits{D.}},
\bauthor{\bsnm{Caleffi}, \binits{M.}},
\bauthor{\bsnm{Krsulich}, \binits{K.}},
\bauthor{\bsnm{Tramonto}, \binits{F.}},
\bauthor{\bsnm{Agliardi}, \binits{G.}},
\bauthor{\bsnm{Prati}, \binits{E.}},
\bauthor{\bsnm{Cacciapuoti}, \binits{A.S.}}:
\batitle{Optimized compiler for distributed quantum computing}.
\bjtitle{ACM T. Quantum Comput.}
\bvolume{4}(\bissue{2}),
\bfpage{15 }
(\byear{2023}).
\doiurl{10.1145/3579367}
\end{barticle}
\endbibitem

\bibitem[\protect\citeauthoryear{Saeedi and Pedram}{2013}]{Saeedi2013}
\begin{barticle}
\bauthor{\bsnm{Saeedi}, \binits{M.}},
\bauthor{\bsnm{Pedram}, \binits{M.}}:
\batitle{Linear-depth quantum circuits for $n$-qubit Toffoli gates with no ancilla}.
\bjtitle{Phys. Rev. A}
\bvolume{87}(\bissue{6}),
\bfpage{062318 }
(\byear{2013}).
\doiurl{10.1103/PhysRevA.87.062318}
\end{barticle}
\endbibitem


\bibitem[\protect\citeauthoryear{Maslov}{2016}]{Maslov2016}
\begin{barticle}
\bauthor{\bsnm{Maslov}, \binits{D.}}:
\batitle{Advantages of using relative-phase Toffoli gates with 
an application to multiple control Toffoli optimization}.
\bjtitle{Phys. Rev. A}
\bvolume{93}(\bissue{2}),
\bfpage{022311 }
(\byear{2016}).
\doiurl{10.1103/PhysRevA.93.022311}
\end{barticle}
\endbibitem

\bibitem[\protect\citeauthoryear{Jun and Choi}{2023}]{Jun2023}
\begin{barticle}
\bauthor{\bsnm{Jun}, \binits{J.-M.}},
\bauthor{\bsnm{Choi}, \binits{I.-C.}}:
\batitle{Optimal multi-bit Toffoli gate synthesis}.
\bjtitle{IEEE Access}
\bvolume{11},
\bfpage{27342}--\blpage{27351}
(\byear{2023}).
\doiurl{10.1109/ACCESS.2023.3243798}
\end{barticle}
\endbibitem

\bibitem[\protect\citeauthoryear{Silva and Park}{2022}]{Silva2022}
\begin{barticle}
\bauthor{\bsnm{da Silva}, \binits{A.J.}},
\bauthor{\bsnm{Park}, \binits{D.K.}}:
\batitle{Linear-depth quantum circuits for multiqubit controlled gates}.
\bjtitle{Phys. Rev. A}
\bvolume{106}(\bissue{4}),
\bfpage{042602 }
(\byear{2022}).
\doiurl{10.1103/PhysRevA.106.042602}
\end{barticle}
\endbibitem

\bibitem[\protect\citeauthoryear{Vale et.~al}{2023}]{Vale2023}
\begin{barticle}
\bauthor{\bsnm{Vale}, \binits{R.}},
\bauthor{\bsnm{Azevedo}, \binits{T.M.D.}},
\bauthor{\bsnm{Ara\'ujo}, \binits{I.C.S.}},
\bauthor{\bsnm{Araujo}, \binits{I.F.}},
\bauthor{\bsnm{da Silva}, \binits{A.J.}}:
\batitle{Circuit decomposition of multicontrolled special unitary single-qubit gates}.
\bjtitle{IEEE Transactions on CAD}
\bvolume{43}(\bissue{3}),
\bfpage{802}--\blpage{811}
(\byear{2023}).
\doiurl{10.1109/TCAD.2023.3327102}
\end{barticle}
\endbibitem

\bibitem[\protect\citeauthoryear{Silva et.~al}{2023}]{Silva2023}
\begin{barticle}
\bauthor{\bsnm{Silva}, \binits{J.D.S.}},
\bauthor{\bsnm{Azevedo}, \binits{T.M.D.}},
\bauthor{\bsnm{Araujo}, \binits{I.F.}},
\bauthor{\bsnm{da Silva}, \binits{A.J.}}:
\batitle{Linear decomposition of approximate multi-controlled single qubit gates}
(\byear{2023}).
\url{https://doi.org/10.48550/arXiv.2310.14974}
\end{barticle}
\endbibitem

\bibitem[\protect\citeauthoryear{He et.~al}{2017}]{He2017}
\begin{barticle}
\bauthor{\bsnm{He}, \binits{Y.}},
\bauthor{\bsnm{Luo}, \binits{M.-X.}},
\bauthor{\bsnm{Zhang}, \binits{E.}},
\bauthor{\bsnm{Wang}, \binits{H.-K.}},
\bauthor{\bsnm{Wang}, \binits{X.-F.}}:
\batitle{Decompositions of n-qubit Toffoli gates with linear circuit complexity}.
\bjtitle{Int. J. Theor. Phys.}
\bvolume{56},
\bfpage{2350}--\blpage{2361}
(\byear{2017}).
\doiurl{10.1007/s10773-017-3389-4}
\end{barticle}
\endbibitem

\bibitem[\protect\citeauthoryear{Balauca and Arusoaie}{2022}]{Balauca2022}
\begin{bchapter}
\bauthor{\bsnm{Balauca}, \binits{S.}},
\bauthor{\bsnm{Arusoaie }, \binits{A.}}:
\batitle{Efficient constructions for simulating multi controlled quantum gates}.
In: \beditor{\bsnm{Groen}, \binits{D.}},
\beditor{\bsnm{de Mulatier}, \binits{C.}},
\beditor{\bsnm{Paszynski}, \binits{M.}},
\beditor{\bsnm{Krzhizhanovskaya}, \binits{V.V.}},
\beditor{\bsnm{Dongarra}, \binits{J.J.}},
\beditor{\bsnm{Sloot}, \binits{P.M.A.}}  (eds.)
\bbtitle{Computational Science - ICCS 2022},
vol. \bseriesno{13353},
pp. \bfpage{179}--\blpage{194}.
\bpublisher{Springer International Publishing},
\blocation{Cham}
(\byear{2022})
\bisbn{978-3-031-08759-2}
\doiurl{https://doi.org/10.1007/978-3-031-08760-8_16}
\end{bchapter}
\endbibitem


\bibitem[\protect\citeauthoryear{Arsoski}{2024}]{Arsoski2024}
\begin{barticle}
\bauthor{\bsnm{Arsoski}, \binits{V.V.}}:
\batitle{Implementing multi-controlled X gates using the quantum Fourier transform}.
\bjtitle{Quantum Inf. Process.}
\bvolume{23},
\bfpage{305 }
(\byear{2024}).
\doiurl{10.1007/s11128-024-04511-w}
\end{barticle}
\endbibitem


\bibitem[\protect\citeauthoryear{Draper}{2000}]{Draper2000}
\begin{barticle}
\bauthor{\bsnm{Draper}, \binits{T.G.}}:
\batitle{Addition on a Quantum Computer}
(\byear{2000}).
\url{https://doi.org/10.48550/arXiv.quant-ph/0008033}
\end{barticle}
\endbibitem


\bibitem[\protect\citeauthoryear{Barenco et.~al}{1996}]{Barenco1996}
\begin{barticle}
\bauthor{\bsnm{Barenco}, \binits{A.}},
\bauthor{\bsnm{Ekert}, \binits{A.}},
\bauthor{\bsnm{Suominen}, \binits{K.-A.}},
\bauthor{\bsnm{T\"{o}rm\"{a}}, \binits{P.}}:
\batitle{Approximate quantum Fourier transform and decoherence}.
\bjtitle{Phys. Rev. A}
\bvolume{54}(\bissue{1}),
\bfpage{139}--\blpage{146}
(\byear{1996}).
\doiurl{10.1103/PhysRevA.54.139}
\end{barticle}
\endbibitem


\bibitem[\protect\citeauthoryear{Maslov}{2007}]{Maslov2007}
\begin{barticle}
\bauthor{\bsnm{Maslov}, \binits{D.}}:
\batitle{Linear depth stabilizer and quantum Fourier 
transformation circuits with no auxiliary qubits
in finite-neighbor quantum architectures}.
\bjtitle{Phys. Rev. A}
\bvolume{76}(\bissue{5}),
\bfpage{052310 }
(\byear{2007}).
\doiurl{10.1103/PhysRevA.76.052310}
\end{barticle}
\endbibitem


\bibitem[\protect\citeauthoryear{Iten}{2016}]{Iten2016}
\begin{barticle}
\bauthor{\bsnm{Iten}, \binits{R.}},
\bauthor{\bsnm{Colbeck}, \binits{R.}},
\bauthor{\bsnm{Kukuljan}, \binits{I.}},
\bauthor{\bsnm{Home}, \binits{J.}},
\bauthor{\bsnm{Christandl}, \binits{M.}}:
\batitle{Quantum circuits for isometries}.
\bjtitle{Phys. Rev. A}
\bvolume{93}(\bissue{5}),
\bfpage{032318 }
(\byear{2016}).
\doiurl{10.1103/PhysRevA.93.032318}
\end{barticle}
\endbibitem


\bibitem[\protect\citeauthoryear{QISKit SDK}{2017}]{Qiskit}
\begin{botherref}
\url{https://pypi.org/project/qiskit/}
\end{botherref}
\endbibitem


\bibitem[\protect\citeauthoryear{qccirc}{2025}]{qccirc}
\begin{botherref}
\url{https://github.com/Arsoski/qccirc.git}
\end{botherref}
\endbibitem


\bibitem[\protect\citeauthoryear{Zheng}{2022}]{Zheng2022}
\begin{barticle}
\bauthor{\bsnm{Zheng}, \binits{W.}},
\bauthor{\bsnm{Zhang}, \binits{Y.}},
\bauthor{\bsnm{Dong}, \binits{Y.}},
\bauthor{\bsnm{Xu}, \binits{J.}},
\bauthor{\bsnm{Wang}, \binits{Z.}},
\bauthor{\bsnm{Wang}, \binits{X.}},
\bauthor{\bsnm{Li}, \binits{Y.}},
\bauthor{\bsnm{Lan}, \binits{D.}},
\bauthor{\bsnm{Zhao}, \binits{J.}},
\bauthor{\bsnm{Li}, \binits{S.}},
\bauthor{\bsnm{Tan}, \binits{X.}},
\bauthor{\bsnm{Yu}, \binits{Y.}}:
\batitle{Optimal control of stimulated Raman adiabatic 
passage in a superconducting qudit}.
\bjtitle{Npj Quantum Inf.}
\bvolume{8}(\bissue{1}),
\bfpage{9 }
(\byear{2022}).
\doiurl{10.1038/s41534-022-00521-7}
\end{barticle}
\endbibitem

\bibitem[\protect\citeauthoryear{Saha}{2022}]{Saha2022}
\begin{barticle}
\bauthor{\bsnm{Saha}, \binits{A.}},
\bauthor{\bsnm{Majumdar}, \binits{D.}},
\bauthor{\bsnm{Saha}, \binits{D.}},
\bauthor{\bsnm{Chakrabarti}, \binits{A.}},
\bauthor{\bsnm{Sur-Kolay}, \binits{S.}}:
\batitle{Asymptotically improved circuit for a $d$-ary 
Grover's algorithm with advanced decomposition of the 
$n$-qudit Toffoli gate}.
\bjtitle{Phys. Rev. A}
\bvolume{105}(\bissue{6}),
\bfpage{062453 }
(\byear{2022}).
\doiurl{10.1103/PhysRevA.105.062453}
\end{barticle}
\endbibitem

\end{thebibliography}
\end{document}